\documentclass[12pt]{article}

\usepackage{kotex}
\usepackage{graphicx}
\usepackage{array}
\usepackage{tabularx}
\usepackage{amsmath,amssymb,bbold}
\numberwithin{equation}{section}
\usepackage{chngcntr}
\usepackage{url}
\usepackage{rotating}
\usepackage{color}
\usepackage{hhline}
\usepackage{hyperref}
\usepackage{slashed}
\usepackage{makecell}
\usepackage{empheq}
\usepackage{physics}
\usepackage{comment}
\usepackage{authblk}
\usepackage{xcolor}
\usepackage[dvips=false,pdftex=false,vtex=false]{geometry}
\usepackage{changepage}
\usepackage{simpler-wick}
\usepackage{makecell}
\usepackage{multirow}
\usepackage{graphicx}
\usepackage{cleveref}
\usepackage{cite}
\usepackage{diagbox}
\usepackage{setspace}
\usepackage{relsize}
\usepackage{soul}
\usepackage{cancel}
\usepackage[column=O]{cellspace}
\newcommand{\sfrac}[2]{\mbox{$\frac{#1}{#2}$}}
\newcommand{\lam}{\lambda}

\newcommand{\be}{\begin{equation}}
\newcommand{\ee}{\end{equation}}
\newcommand{\bea}{\begin{eqnarray}}
\newcommand{\eea}{\end{eqnarray}}
\newcommand{\eq}[1]{Eq.~(\ref{#1})}

\begin{document}

\newcommand{\s}{\\ \vspace*{-2mm}}

\begin{flushright}
KIAS-Q25016
\end{flushright}

\vskip 1.cm

\begin{center}
{\Large \bf Causal Bounds on EFTs with anomalies 
\vskip .4cm
with  a Pseudoscalar,
 Photons, and Gravitons}\\[20pt]
{Ziyu Dong${}^{a,}$\footnote{zdong@ictp.it},
Jaehoon Jeong${}^{b,}$\footnote{jeong229@kias.re.kr, jeong229@gmail.com} and Alex Pomarol${}^{c,d,}$\footnote{alex.pomarol@uab.cat}} \\[15pt]
${}^a${\it  The Abdus Salam ICTP, Strada Costiera 11, 34135, Trieste, Italy}\\
${}^b${\it  Quantum Universe Center, KIAS, Seoul 02455, Korea}\\
${}^c${\it IFAE and BIST, Universitat Aut\`onoma de Barcelona, 08193 Bellaterra, Barcelona}\\
${}^d${\it Departament de F\'isica, Universitat Aut\`onoma de Barcelona, 08193 Bellaterra, Barcelona}\\

\end{center}

\vskip 0.5cm

%
%

\begin{abstract}
Theories with pseudoscalars that couple through anomalies (such as axion models) are of particular phenomenological interest. We carry out a comprehensive analysis of all bounds obtainable from bootstrapping the amplitudes when a pseudoscalar couples to photons and gravitons. 
This allows us to find  new  cutoff scales of  theories with anomalies that are  more restrictive than those obtained from naive perturbative analysis. Our results are especially relevant for holographic models, as the bounds determine the allowed region of the five-dimensional EFTs,  for example, by imposing strong bounds on Chern-Simons terms.  
We also consider modifications of General Relativity in photon--graviton couplings and show that current experiments are sensitive to these effects only if new physics appears at $\sim 10^{-10}$ eV.

\end{abstract}

\tableofcontents

\section{Introduction}

The $S$-matrix bootstrap program, which consists of imposing fundamental principles such as unitarity, causality, and crossing symmetry on scattering amplitudes, has proven to be a very powerful tool for constraining Effective Field Theories (EFTs)
\cite{Kruczenski:2022lot,
Paulos:2016but,
Paulos:2017fhb,
Guerrieri:2018uew,
Doroud:2018szp,
deRham:2018qqo,
Zhang:2018shp,
Bellazzini:2019bzh,
Remmen:2019cyz,
Bellazzini:2019xts,
Huang:2020nqy,
Sinha:2020win,
Bellazzini:2020cot,
Tolley:2020gtv,
Hebbar:2020ukp,
Arkani-Hamed:2020blm,
Caron-Huot:2020cmc,
Caron-Huot:2021rmr,
Bern:2021ppb,
Chiang:2021ziz,
Henriksson:2021ymi,
Zahed:2021fkp,
Alvarez:2021kpq,
EliasMiro:2022xaa,
Serra:2022pzl,
Caron-Huot:2022ugt,
Caron-Huot:2022jli,
Henriksson:2022oeu,
Albert:2022oes,
Fernandez:2022kzi,
Acanfora:2023axz,
He:2023lyy,
Guerrieri:2023qbg,
Albert:2023jtd,
Ma:2023vgc,
Bellazzini:2023nqj,
Li:2023qzs,
Albert:2023seb,
Hong:2023zgm,
He:2024nwd,
Guerrieri:2024jkn,
Beadle:2024hqg,
Bertucci:2024qzt,
Albert:2024yap,
Xu:2024iao,
Berman:2024kdh,
Haring:2024wyz,
Dong:2024omo,
Beadle:2025cdx,
Chang:2025cxc,
Pasiecznik:2025eqc,
Huang:2025icl
}.
By relying only on general consistency conditions, this approach has provided robust, model-independent constraints on the (Wilson) coefficients of higher-dimensional operators.

Theories with anomalies are of special interest in this context, as certain higher-dimensional operators, such as those describing the coupling of a Goldstone boson to photons or gravitons, are entirely fixed by symmetry considerations. 
In such cases, bootstrap techniques can be used to determine new cutoff scales for the theories, which are more stringent than naive perturbative estimates \cite{Serra:2022pzl,Dong:2024omo}. 
In the present work, we build upon the analysis of Ref.~\cite{Dong:2024omo} and undertake a systematic study of all bounds that can be derived from bootstrapping amplitudes involving pseudoscalars, photons, and gravitons.

The inclusion of gravitons plays a central role. 
The spin-2 nature of the graviton causes scattering amplitudes to grow more rapidly with energy, making the leading Wilson coefficients directly accessible through dispersion relations. 
Moreover, gravity provides a double-subtracted dispersion relation whose positivity allows one to constrain higher-dimensional operator coefficients as a function of the Newton constant $G_N$. 
The price to pay is that dispersion relations must be smeared in order to obtain nontrivial bounds \cite{Caron-Huot:2021rmr,Caron-Huot:2022ugt,Caron-Huot:2022jli}.
Nevertheless, the smearing method has already been proven to be very efficient in many processes (see for example
\cite{Henriksson:2022oeu,Haring:2024wyz,Chang:2025cxc,Beadle:2024hqg,Hong:2023zgm,Xu:2024iao,Pasiecznik:2025eqc,Beadle:2025cdx,Bellazzini:2019xts,Albert:2024yap}).
Alternatively, one could use a time-delay analysis \cite{Camanho:2014apa}, 
where similar bounds could also be derived \cite{Serra:2022pzl}.

We will show that in theories with anomalies, such as axion models, 
causality imposes a cutoff scale which lies below that determined by perturbativity.
 At this cutoff scale,   massive states with spin $J \geq 2$ must necessarily appear. In particular, we will derive
the bound
\be
M_{J\geq 2}\lesssim \sqrt{\frac{c_{4\gamma}}{G_N}} \frac{F_\pi^2 }{\kappa\kappa_g}\,,
\ee
where $\kappa$ ($\kappa_g$) is the coupling of the pseudoscalar to photons (gravitons) dictated by  anomalies,
$F_\pi$ is the decay constant and  $c_{4\gamma}$ is the 4-photon coupling  at low-energies, $ c_{4\gamma} e^4F^4_{\mu\nu}$.
For completeness, we will also study  bounds on other non-minimal couplings.
For example, we will show that the coupling $\kappa_1$ between photons and gravitons,
arising from the $R_{\mu\nu\rho\sigma}F^{\mu\nu}F^{\rho\sigma}$ interaction, 
is related to the presence of $J\geq 3$ states:
 \be
e^2|\kappa_1|\lesssim  \frac{1}{M^2_{J\geq 3}}\,,
\ee
where $e$ is the electric charge.

We will show that  our bounds, although derived in four dimensions (4D), can be applied to constrain 5D theories
with Chern–Simons terms.
They may thus have significant implications for 5D holographic duals of strongly-coupled theories \cite{Aharony:1999ti}.
For example, the Chern–Simons terms can play an important role in
holographic duals of hydrodynamical systems at finite temperature and chemical potential~\cite{Cremonini:2009sy,Delsate:2011qp,Das:2005za,Donos:2012wi,Cai:2012mg,Megias:2013joa,Azeyanagi:2015gqa,Bhattacharyya:2016knk,Liu:2016hqb,Baggioli:2024zfq}, 
where sizeable Chern–Simons coefficients have been shown to  lead to phase transitions, instabilities, or causality violations\cite{Donos:2012wi,Cai:2012mg,Bhattacharyya:2016knk,Liu:2016hqb}.

Furthermore, our results can be straightforwardly generalized to theories involving scalar fields instead of pseudoscalars \cite{Serra:2022pzl}, although in that case the relevant couplings are not fixed by anomalies.

The paper is organized as follows. In Sec.~\ref{sec:3pt}, we introduce all possible 3-point couplings involving pseudoscalars, photons and gravitons ($\eta,\gamma,h$),  and their relation
to anomalies. In Sec.~\ref{sec:bootstrapping_L}, 
we derive the needed dispersion relations and discuss which $2 \to 2$ amplitudes are relevant to bootstrap in order to get bounds on anomalous and non-minimal couplings.
In Sec.~\ref{sec:bound}, we explicitly derive the bounds on $\kappa\kappa_g$ and $\kappa_1$.  
In Sec.~\ref{sec:pheno}, we discuss the implications of the bounds in axion models, strongly-coupled theories, and 5D (holographic) models,  and conclude in Sec.~\ref{sec:conclusion}. In three Appendices we present technical details on the
derivations of the bounds. In Appendix~\ref{appendix:dispersion_relation_at_fixed_w}, we introduce a novel way to derive improved dispersion relations for $t\leftrightarrow u$ symmetric amplitudes. 
In Appendix~\ref{appendix:explicit_smeared}, we list the explicit forms of smeared Wigner $d$-functions used in the main text. In Appendix~\ref{appendix:positivity_of_smeared_d}, we discuss all details for the positivity of smeared Wigner $d$-functions.

\section{Three-point interactions involving $\eta$, $\gamma$, and $h$}
\label{sec:3pt}

Let us begin by examining all possible on-shell 3-point (3-pt) interactions in theories containing a pseudoscalar $\eta$, a photon $\gamma$ and a graviton $h$, which will define the EFT ${\cal L}_{\rm EFT}(\eta,\gamma,h)$ that we want to bootstrap.
We will use the spinor-helicity formalism~\cite{Dixon:2013uaa} with the convention that all states are incoming.  In this work, we restrict for simplicity  to a CP-symmetric EFT.

We first construct the on-shell 3-pt amplitudes characterizing the anomalous couplings of a pseudoscalar (dashed line) to two photons (curly lines) and to two gravitons (double-curly lines). These are given by
\begin{equation}
\raisebox{-0.5\height}{\includegraphics[width=0.9\textwidth]{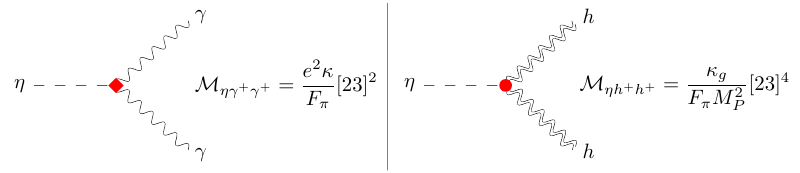}}
\label{eq:3pt_anomaly}
\end{equation}
where $M_P$ denotes the Planck scale, $F_\pi$ the decay constant of the pseudoscalar $\eta$, $e$ the electric coupling, while $\kappa, \kappa_g$ are dimensionless anomaly coefficients. The parametrization is chosen such that each graviton contributes a factor of $1/M_P$, each photon a factor of $e$, and each pseudoscalar a factor of $1/F_\pi$ in the amplitudes. As shown in Eq.~\eqref{eq:3pt_anomaly}, we will mark with 
a red diamond (circle) the  vertex involving  $\kappa$ ($\kappa_g$).
The amplitudes in Eq.~\eqref{eq:3pt_anomaly}  arise in theories with an anomalous $U(1)_\eta$ symmetry, under which  $\eta\to \eta+\theta$. They come from  the following terms in the Lagrangian:
\begin{align}
\frac{i e^2\kappa}{F_\pi}\, \eta \,
\, \varepsilon^{\mu\nu\rho\sigma}
F_{\mu\nu}
F_{\rho\sigma}
\quad \mbox{and}\quad 
\frac{i\kappa_g}{F_\pi}\,
\eta\, \varepsilon^{\mu\nu\rho\sigma}
R_{\mu\nu\alpha\beta}
R_{\rho\sigma}^{\;\;\,\alpha\beta}\,,
\end{align}
where $F_{\mu\nu}$ and $R_{\mu\nu\rho\sigma}$ are the field strength and Riemann curvature tensors.
Therefore $\kappa$ characterizes  the $U(1)_\eta\times U(1)_{\rm EM}\times U(1)_{\rm EM}$ anomaly, while 
$\kappa_g$ corresponds to the $U(1)_\eta$-gravitational anomaly.

Additionally, there are 3-pt amplitudes involving  non-minimal couplings of a graviton to two photons and three gravitons:
\begin{equation}
\raisebox{-0.5\height}{\includegraphics[width=1.02\textwidth]{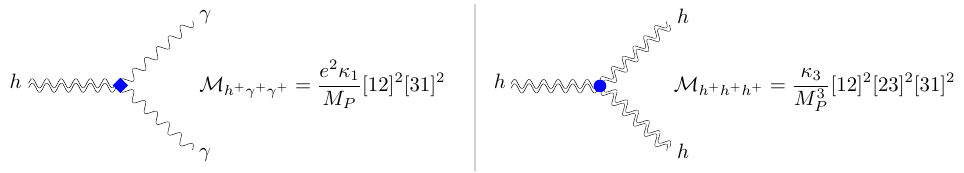}}
\label{eq:3pt_non-minimal}
\end{equation}
where the coefficients $\kappa_1,\kappa_3$ have mass dimension $-2$.
These amplitudes arise respectively from the following higher-dimensional operators:
\begin{align}
e^2\kappa_1\,
R_{\mu\nu\rho\sigma}F^{\mu\nu}F^{\rho\sigma}
\quad \mbox{and}\quad 
\kappa_3\, R_{\mu\nu\rho\sigma}R^{\rho\sigma\alpha\beta}R_{\alpha\beta}^{\;\;\,\mu\nu}\,.
\label{eq:CPeven_Lagrangian}
\end{align}
Apart from the 3-pt amplitudes discussed above, the EFT also contains   
  the minimal gravitational  couplings  
  of a pseudoscalar, a photon, and a graviton, as described in General Relativity (GR).
They correspond to the interactions $h^\pm\eta\eta$, $h^\pm\gamma^+\gamma^-$ and $h^\pm h^+h^-$.

\begin{table}[t]
\renewcommand{\arraystretch}{1.3}
\centering
\begin{tabular}{|c|c|}
 \hline
$h^\pm h^\pm \gamma^+$ & Bose symmetry
\\ \hline
$h^+h^-\gamma^+$ & Factorization consistency
\\ \hline
$h^+\gamma^+\eta \;\&\; h^+\gamma^-\eta$ & Factorization consistency
\\ \hline
$\gamma^+\gamma^+\gamma^+ \;\&\; \gamma^+\gamma^+\gamma^-$ & Bose symmetry
\\ \hline
$\gamma^+\eta\eta$ & Bose symmetry
\\ \hline
$\eta\eta\eta$ & CP symmetry
\\ \hline
\end{tabular}
\caption{Forbidden 3-pt on-shell interactions involving $\eta$, $\gamma$, and $h$.}
\label{tab:forbidden_3pt}
\end{table}
Other 3-pt interactions are forbidden for the reasons shown in Table~\ref{tab:forbidden_3pt}. In particular, the amplitudes with two identical particles are antisymmetric under spinor exchange, violating Bose symmetry;
CP symmetry excludes the interaction of three pseudoscalars, 
and the remaining cases do not generate any 4-pt amplitude 
consistent with factorization \cite{Dixon:2013uaa}.

\section{Bootstrapping    ${\cal L}_{\rm EFT}(\eta,\gamma,h)$  from four-point amplitudes}
\label{sec:bootstrapping_L}

Our goal is to provide an analysis of all possible 4-pt amplitudes 
involving $\eta$, $\gamma$, and $h$ that can yield bounds on anomalous and non-minimal couplings,
\eq{eq:3pt_anomaly} and  \eq{eq:3pt_non-minimal} respectively.
We begin by reviewing  how  analyticity, crossing symmetry, and unitarity impose constraints on the low-energy parameters of scattering amplitudes.
We will consider that the UV completions of ${\cal L}_{\rm EFT}(\eta,\gamma,h)$ 
 consists of theories with (infinite) weakly-coupled  particles,
such that  4-pt amplitudes are  mediated by the exchange of single states.
This makes the analysis less cumbersome. 
Nevertheless, our results also extend to the case where this assumption is relaxed.\footnote{For example, 
new physics appearing in the amplitudes at the loop level were considered  in \cite{Bellazzini:2021shn}, 
where it was  shown that causal   bounds persist.}
Effects of gravity loops have been considered in \cite{Beadle:2025cdx,
Chang:2025cxc}.

\subsection{Smeared dispersion relations}
\label{sec:dispersion}

\begin{figure}[t!]
\centering
\includegraphics[scale=0.85]{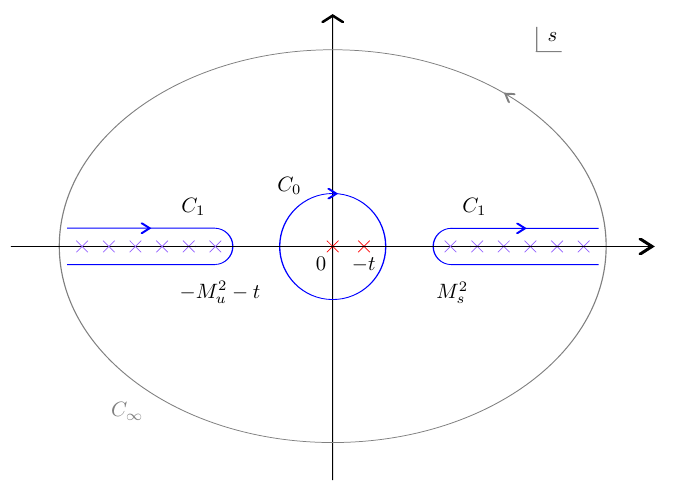}
\caption{Contours in the complex $s$-plane 
used to obtain dispersion relations. 
The red crosses  mark the poles from massless exchanges in the amplitudes at low-energies,
while the purple crosses indicate  poles associated with the heavy states of the UV completion.
}
\label{fig:contour}
\end{figure}

Let us  consider  $2 \to 2$ processes whose amplitudes ${\cal M}$
can be expressed as functions of the Mandelstam 
variables $s,t,u$.
For  fixed real $t<0$,  causality ensures that $\mathcal{M}(s,u)$ is analytic throughout the complex $s$-plane, except along the real axis.
In the case of amplitudes mediated by tree-level exchange,  the singularities reduce to simple poles only on the real $s$-axis, as depicted in Fig.~\ref{fig:contour} where $M_s$ ($M_u$) corresponds to the mass of the first UV state in the $s$-channel ($u$-channel).

We can derived what is called a  $k$-subtracted dispersion relation in the following way. 
Let us consider the  function 
\be
\frac{{\cal M}(s,u)}{s^{1+k-l}(s+t)^{l}}\,,
\ee
where $k$ and $l\leq k$ are nonnegative integers to be specified later. 
By Cauchy’s theorem we have that the integration of this function along the $C_\infty$ contour
of  Fig.~\ref{fig:contour},   $\int_{C_\infty}{\cal M}/(s^{1+k-l}(s+t)^{l})\equiv 2 i R_\infty$, is equivalent, by deforming the contour, 
to   the integral along    $C_0$ and $C_1$ of  Fig.~\ref{fig:contour}.
The contour integral along $C_1$ picks up the non-analytical part of the amplitude that is related to its
imaginary part. This provides  the following relation:
\begin{align}
\hspace{-0.7cm}
\frac{1}{2i}\oint_{C_0}ds' 
\frac{\mathcal{M}(s',-s'-t)}{s^{\prime 1+k-l}(s'+t)^{l}}
=\int^{\infty}_{M_s^2}\!ds'
\frac{\mbox{Im}\mathcal{M}(s',-t-s')}{ s^{\prime 1+k-l}(s'+t)^{l}}
+(-1)^{k}\!\! \int^{\infty}_{M_u^2}\!
ds' \frac{\mbox{Im}\mathcal{M}(-t-s',s')}{(s'+t)^{1+k-l}
s^{\prime l}}+R_\infty(t)\,.
\label{eq:dispersion}
\end{align}
We can expand  ${\rm Im}\mathcal{M}$  in partial-waves~\cite{Jacob:1959at}  using the Wigner $d$-functions. In particular,
\begin{align}
\mbox{Im}\mathcal{M}_{\lambda_1,\lambda_2,\lambda_3,\lambda_4}(s,-t-s)
&=\sum_J(2J+1)\rho_J(s)d^J_{\lam_{12},\lam_{43}}\bigg(1+\frac{2t}{s}\bigg)\,,
\nonumber
\\
\mbox{with}\quad (2J+1)\rho_J(s)&=
\pi\sum_i g_{\lam_1\lam_2,i}\,g_{\lam_3\lam_4,i}\; m_i^2 \delta(s-m_i^2)
\,\delta_{J,J_i}\,,
\label{scha}
\end{align}
where $\lambda_{ab}=\lambda_{a}-\lambda_{b}$ with  $\lambda_a$ being the helicity of the external state $a$, and the couplings $g_{\lambda_a\lambda_b,i}^2$ appear from the residues of the $s$-channel poles.  The corresponding $u$-channel expression follows from \eq{scha} with the replacement $s \rightarrow u$ together with $\lambda_{2}\leftrightarrow \lambda_{4}$.
We also introduce, for convenience, the ``high-energy average"
\begin{align}
\langle (...)\rangle
=\frac{1}{\pi}
\int^\infty_{M^2_{s,u}} \frac{dm^2}{m^2}
\sum_J (2J+1)\rho^J(m^2)(...)\,.
\end{align}
With all this, Eq.~\eqref{eq:dispersion} can be recast in the following way:
\begin{align}
\hspace{-0.4cm}
\frac{1}{2\pi i}\oint_{C_0}ds' 
\frac{\mathcal{M}(s',-s'-t)}{s^{\prime 1+k-l}(s'+t)^{l}}
=\bigg\langle 
\frac{d^J_{\lambda_{12},\lambda_{43}}(1+2t/m^2)}{m^{2(k-l)}(m^2+t)^{l}}
\bigg \rangle
+(-1)^{k} \bigg\langle 
\frac{ d^J_{\lambda_{14},\lambda_{23}}(1+2t/m^2)}{(m^2+t)^{1+k-l}m^{2l-2}}
\bigg \rangle+R_\infty(t)\,.
\label{eq:dispersion2}
\end{align}
The dependence  on  $R_\infty$ makes the above relation no very useful. In principle, the
Froissart-Martin bound~\cite{Froissart:1961ux,Martin:1965jj} tells us that for  $k\geq 2$,  $R_\infty\to 0$. Nevertheless, we are dealing with $\gamma$ and
$h$ as internal massless states, and therefore the Froissart-Martin bound is not directly applicable.
To get rid of $R_\infty$ in \eq{eq:dispersion2}, we will integrate this relation over $t$ with a proper ``smearing function" $A(t)$ to be discussed later.  The reason for this smearing is twofold. On one hand, we avoid singularities $1/t$ due to the graviton pole in the amplitudes. On the other hand, it can be proven \cite{Caron-Huot:2022ugt}  that  $\int^0_{-|t|_{\rm max}}\!\!\! dt\, A(t)\, R_\infty(t)\to 0$ for $k\geq 2$. 
In this way, we get the smeared dispersion relation:
\begin{align}
\hspace{-0.2cm}
\frac{1}{2\pi i}\!\!\int^0_{-|t|_{\rm max}} \!\!\!\!\!\!\!dt\, A(t)\!\!\oint_{C_0}\!\!ds' 
\frac{\mathcal{M}(s',-s'-t)}{s^{\prime 1+k-l}(s'+t)^{l}}
=\!\!\int^0_{-|t|_{\rm max}} \!\!\!\!\!\!\! dt\, A(t)\!\!\left[\!\bigg\langle\! 
\frac{d^J_{\lambda_{12},\lambda_{43}}(1+\frac{2t}{m^2})}{m^{2(k-l)}(m^2+t)^{l}}
\!\bigg \rangle
\!+\!(-1)^{k} \bigg\langle 
\!\frac{ d^J_{\lambda_{14},\lambda_{23}}(1+\frac{2t}{m^2})}{(m^2+t)^{1+k-l}m^{2l-2}}
\!\bigg \rangle\!\right].
\label{eq:dispersion3}
\end{align}

\subsection{Relevant four-point amplitudes with  $\eta$, $\gamma$ and $h$}
\label{sec:4pt-tree}

\newcolumntype{M}[1]{>{\centering\arraybackslash}m{#1}}
\begin{table}[t]
\renewcommand{\arraystretch}{1.45}
\centering
\resizebox{0.9\textwidth}{!}{
\begin{tabular}{|c||c||c|c||c||c|c|}
\hline
&\multicolumn{3}{c||}{Bound derived}      & \multicolumn{3}{c|}{No bound derived} \\
\hline
\multirow{6}{*}[1.15cm]{Ref.~\cite{Dong:2024omo}}  
&\multirow{2}{*}[0.35cm]{$h^+h^+h^-h^-$} 
&\multirow{2}{*}[0.35cm]{$\kappa_g^2\;\&\; \kappa_3^2$}
&\multirow{2}{*}[0.35cm]{$k=3$}
&\multirow{3}{*}[0.53cm]{$h^+h^+h^+h^+$} 
&\multirow{3}{*}[0.53cm]{$\kappa_g^2\;\&\; \kappa_3$}
&\multirow{3}{*}[0.53cm]{$k=2$}
\\[-10.5pt]
&
& 
&
&
&
&
\\[-10.5pt]
\cline{2-4}
&\multirow{2}{*}[0.33cm]{$h^+h^-\eta\eta$}
&\multirow{2}{*}[0.33cm]{$\kappa_g^2$}
&\multirow{2}{*}[0.33cm]{$k=2$}
& 
&
&
\\[-10pt]
\cline{5-7}
&
&
&
&\multirow{3}{*}[0.53cm]{$\eta\eta\eta h^+$} 
&\multirow{3}{*}[0.53cm]{$\kappa_g$}
&\multirow{3}{*}[0.53cm]{$k=2$}
\\[-10.5pt]
\cline{2-4}
&\multirow{2}{*}[0.35cm]{$h^+h^+h^-\eta$}
&\multirow{2}{*}[0.35cm]{$\kappa_g\;\&\; \kappa_g\kappa_3$}
&\multirow{2}{*}[0.35cm]{$k=2$}
& 
&
&
\\[-10.5pt]
&
& \\[-10.5pt]
\hline \hline
\multirow{4}{*}{This work}  
&$h^+h^+\gamma^-\gamma^-$
&$\kappa\kappa_g \;\&\; \kappa_1 \kappa_3$
&$k=2$
&\multirow{2}{*}{$\gamma^+\gamma^- h^+\eta$} 
&\multirow{2}{*}{$\kappa_g \;\&\; \kappa\kappa_1$}
&\multirow{2}{*}{$k=2$}
\\
\cline{2-4}
&$h^+h^+\gamma^+\gamma^+$
&$\kappa \kappa_g \;\&\; \kappa_1$
&$k=2$
& 
&
&
\\
\cline{2-7}
&$h^+h^-\gamma^+\gamma^+$
&$\kappa_1$
&$k=2$
& \multirow{2}{*}{$\gamma^+\gamma^+ h^-\eta$}
& \multirow{2}{*}{$\kappa_1\kappa_g$}
& \multirow{2}{*}{$k=2$}
\\
\cline{2-4}
&$h^+h^-\gamma^+\gamma^-$
&$\kappa_1^2$
&$k=3$
& 
&
&
\\
\hline
\end{tabular}
}
\caption{External states, $\eta$, $\gamma^{\pm}$ and $h^\pm$, of an amplitude whose  $k$-subtracted dispersion relation    involves the anomaly coefficients ($\kappa$, $\kappa_g$) or  non-minimal  couplings ($\kappa_1$, $\kappa_3$). 
The first line shows the amplitudes studied in Ref.~\cite{Dong:2024omo}, while the second are those discussed in the present work. We also indicate whether a  bound on the 3-pt coefficients can be obtained. Amplitudes  not shown do not involve the 3-pt couplings of  \eq{eq:3pt_anomaly} and \eq{eq:3pt_non-minimal}.}
\label{tab:anomaly_contents}
\end{table}

Armed with the smeared dispersion relations \eq{eq:dispersion3}, we aim to obtain all possible bounds on the 
couplings of \eq{eq:3pt_anomaly} and \eq{eq:3pt_non-minimal} using 
all amplitudes involving as external states $\eta$, $\gamma$ and $h$.  
In Table~\ref{tab:anomaly_contents}  
we present a summary of the  
amplitudes whose $k$-subtracted dispersion relation involves any of the couplings
of \eq{eq:3pt_anomaly} and \eq{eq:3pt_non-minimal}.
The table also specifies whether a bound can be obtained or not.
A brief comment is in order.

Amplitudes involving only $\eta$ and $h$ as external states were already studied in Refs.~\cite{Dong:2024omo,Serra:2022pzl}, and bounds on  $\kappa_g$ and $\kappa_3$ were obtained there.
Therefore  we will not discuss them here any longer.\footnote{The presence of $\gamma$ as an internal state does not  afffect  the dispersion relations of these amplitudes.}

We will center in  processes involving  $\gamma$. 
In particular, we will consider  $\gamma^+\gamma^+\rightarrow h^+h^+$ that is sensitive to  $\kappa\kappa_g $ and $\kappa_1\kappa_3$, as shown  in Fig.~\ref{fig:4pt_kappa}.
From this process we will be able to derive a bound on $\kappa\kappa_g$.
Similarly,   the process $\gamma^+\gamma^+\rightarrow h^-h^-$  also involve
$\kappa\kappa_g$ and a bound will also be derived, although it will be more model dependent.

Additionally, we will consider the process $\gamma^+h^+\rightarrow \gamma^-h^+$, shown in  Fig.~\ref{fig:4pt_kappa1}, which is sensitive to  $\kappa_1$ thanks to the    
$\gamma$ and $h$ exchange, that will allow to bound this coupling. 
The elastic process $\gamma^+h^-\rightarrow \gamma^+h^-$ could also be used
to bound $\kappa_1$,   although it leads to a similar bound, and  we will  not derive it here.

In Table~\ref{tab:anomaly_contents} we also show other 4-pt amplitudes from which, although their dispersion relations involve anomalous or non-minimal couplings, we have not been able to derive any bound. This is because we always encounter infinitely many Wilson coefficients in these cases.
Other amplitudes not shown 
in Table~\ref{tab:anomaly_contents} do not involve the
3-pt couplings of  \eq{eq:3pt_anomaly} and \eq{eq:3pt_non-minimal}.

\section{Bootstrapping $\kappa\kappa_g$ and $\kappa_{1}$}
\label{sec:bound}

\subsection{$\gamma^+\gamma^+\rightarrow h^+ h^+$}
\label{sec:bound_rrhh1}

\begin{figure}[t]
\centering
\includegraphics[scale=1.1]{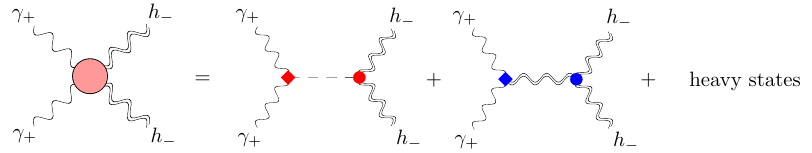}
\caption{Contributions to the process $\gamma^+\gamma^+\rightarrow h^+h^+$. All helicities are indicated in the all-incoming convention. The red diamond and circle represent resprectively the anomaly couplings $\kappa$ and $\kappa_g$, while the blue diamond and circle correspond to the non-minimal couplings $\kappa_1$ and $\kappa_3$. 
}
\label{fig:4pt_kappa}
\end{figure}

At low-energies the amplitude for the scattering $\gamma^+\gamma^+\rightarrow h^+ h^+$ is given by 
(see Fig.~\ref{fig:4pt_kappa})
\begin{align}
&\mathcal{M}_{\gamma^+\gamma^+h^-h^-}=
s^3\,
\bigg(\frac{e^2\kappa \kappa_g}{F_\pi^2 M_P^2 s}
+
\frac{e^2\kappa_{1}\kappa_3 tu}{ M_P^4s}
+\cdots
\bigg),
\label{eq:EFT_rrhh}
\end{align}
where  $\kappa \kappa_g$ appears
from the $s$-channel $\eta$  exchange, while $\kappa_1\kappa_3$  from the $h$ exchange.\footnote{We omit a little-group helicity phase, which has no impact on analyticity.} 
Here the symbol $\cdots$ corresponds  to contact-term polynomials in  $s,t,u$.

\begin{figure}[t!]
\centering
\includegraphics[scale=1]{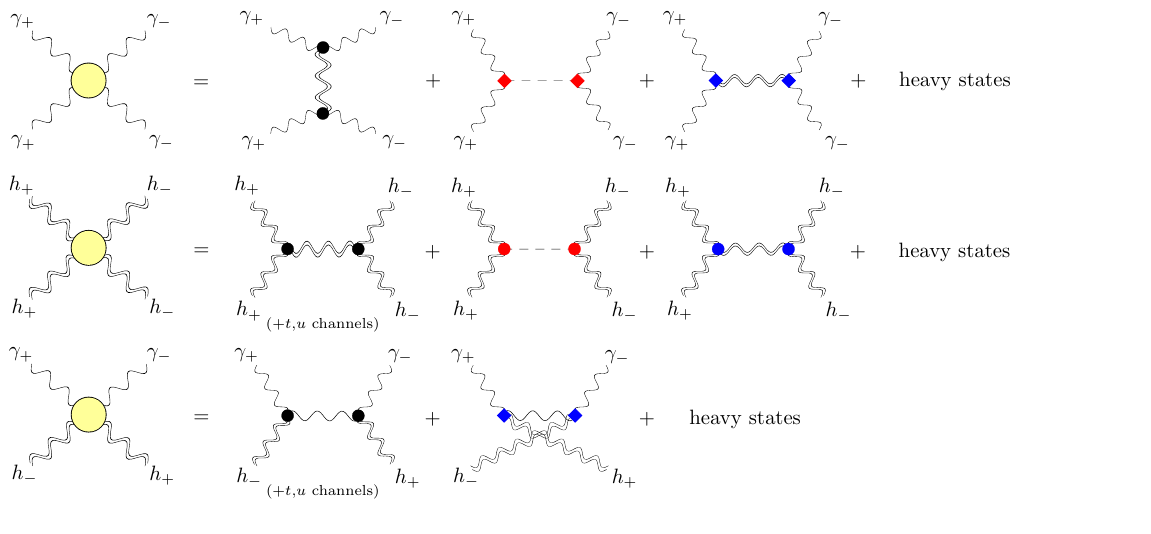}
\caption{Contributions to  the three elastic processes 
$\gamma^+\gamma^+\rightarrow \gamma^+\gamma^+$, 
$h^+h^+\rightarrow h^+h^+$, 
and $\gamma^+h^- \rightarrow \gamma^+ h^-$. 
The red and blue markers correspond to the couplings introduced in Fig.~\ref{fig:4pt_kappa}. 
The black circle represents the gravitational minimal coupling.}
\label{fig:4pt_for_bounds}
\end{figure}

Because \eq{eq:EFT_rrhh} is an inelastic amplitude, for which positivity bounds do not hold, 
we must relate it to elastic amplitudes:  
$\gamma^+\gamma^+\rightarrow\gamma^+\gamma^+$, $h^+h^+\rightarrow h^+h^+$, and $\gamma^+h^-\rightarrow\gamma^+ h^-$, shown in Fig.~\ref{fig:4pt_for_bounds}.
At low-energies these are given  by
\begin{align}
&\mathcal{M}_{\gamma^+\gamma^+\gamma^-\gamma^-}=
s^2\,
\bigg(\frac{1}{M_P^2 w}
+
\frac{e^4\kappa_{1}^2 w}{M_P^2}
-\frac{e^4\kappa^2}{F_\pi^2s}
+e^4c_{4\gamma}
+\cdots
\bigg),\ \ \ w\equiv\frac{tu}{s},
\label{eq:EFT_rrrr}
\\
&\mathcal{M}_{h^+h^+h^-h^-}=
s^4\,
\bigg(
\frac{1}{M_P^2 stu}
-
\frac{\kappa^2_g}{F_\pi^2 M_P^2 s}
+
\frac{\kappa_3^2 tu}{M_P^6 s}
+\cdots
\bigg),
\label{eq:EFT_hhhh}
\\
&\mathcal{M}_{\gamma^+h^-\gamma^-h^+}=-
s u^3 \,
\bigg(
\frac{1}{M_P^2 stu}
-
\frac{e^4\kappa_{1}^2}{M_P^2 u}
+\cdots
\bigg)\,.
\label{eq:EFT_rhrh}
\end{align}
For the four-photon amplitude we have introduced a new variable, $w=tu/s$. 
The reason is the following. $k=2$ dispersion relations at fixed $t$ involve 
infinitely many low-energy Wilson coefficients from contact terms, and are therefore impractical.
Nevertheless, taking the amplitude at fixed $w$, we will be able to 
obtain dispersion relations that  only depend on one   Wilson coefficient, $c_{4\gamma}$
(see Appendix.~\ref{appendix:dispersion_relation_at_fixed_w} for details). That is, in the case of $t\leftrightarrow u$ symmetry, the variable substitution of $w$ can directly obtain the improved dispersion relation in \cite{Caron-Huot:2021rmr,Caron-Huot:2022ugt}.

Let us derive $k=2$  dispersion relations  for these  four processes.
Using  \eq{eq:dispersion2} with $l=0$ and  \eq{eq:EFT_rrhh} for the inelastic process,
 we get
\begin{align}
\frac{e^2\kappa \kappa_g}{F_\pi^2 M_P^2}
-
\frac{e^2\kappa_{1}\kappa_3 t^2}{ M_P^4}
&=
\bigg\langle
\frac{d^J_{0,0}\big(1+\frac{2t}{m^2}\big)}{m^4}
\bigg \rangle^{g_{\gamma^+\gamma^+}}_{g_{h^-h^-}}
+
\bigg\langle
\frac{m^2(-1)^J d^J_{3,3}\big(1+\frac{2t}{m^2}\big)}{(m^2+t)^3}
\bigg\rangle^{g_{\gamma^+h^-}}_{g_{\gamma^+h^-}}+R_\infty (t)\,.
\label{eq:dispersion_rrhh}
\end{align}
For the elastic processes, we have using \eq{eq:dispersion_rel_at_fixed_w}
with \eq{eq:EFT_rrrr}:
\be
\frac{1}{M_P^2 w}+\frac{e^4\kappa_1^2 w}{M_P^2}+e^4c_{4\gamma}
=
\Bigg\langle
\frac{d^J_{0,0}\Big(\sqrt{1-\frac{4w}{m^2}}\Big)}{m^4}
\Bigg\rangle^{g_{\gamma^+\gamma^+}}_{g_{\gamma^-\gamma^-}}
+
\Bigg\langle
\frac{d^J_{2,2}\Big(\frac{m^2-w}{m^2+w}\Big)}{m^8}(m^2+2w)(m^2+w)
\Bigg\rangle^{g_{\gamma^+\gamma^-}}_{g_{\gamma^-\gamma^+}}+R_\infty(t)\,,
\ee
and from \eq{eq:dispersion2}  with $l=0$ and $2$, together 
with \eq{eq:EFT_hhhh} and \eq{eq:EFT_rhrh} respectively, we get
\begin{align}
-\frac{1}{M_P^2 t}
&=
\bigg\langle
\frac{d^J_{0,0}\big(1+\frac{2t}{m^2}\big)}{m^4}
\bigg \rangle^{g_{h^+h^+}}_{g_{h^- h^-}}
+
\bigg\langle
\frac{m^2 d^J_{4,4}\big(1+\frac{2t}{m^2}\big)}{(m^2+t)^3}
\bigg\rangle^{g_{h^+h^-}}_{g_{h^- h^+}}+R_\infty(t),
\\
-\frac{1}{M_P^2 t}
&=
\bigg\langle
\frac{d^J_{3,3}\big(1+\frac{2t}{m^2}\big)}{(m^2+t)^2}
\bigg \rangle^{g_{\gamma^+h^-}}_{g_{\gamma^-h^+}}
+
\bigg\langle
\frac{m^2 d^J_{1,1}\big(1+\frac{2t}{m^2}\big)}{m^2(m^2+t)}
\bigg \rangle^{g_{\gamma^+h^+}}_{g_{\gamma^-h^-}}+R_\infty(t).
\label{eq:dispersion_rhrh}
\end{align}

\noindent
The super- and subscripts denote the couplings appearing in the residues of the $s$- and $u$-channel poles, respectively.   As we commented before,
smearing over these relations with smearing functions, $A(t),B(t),...$, we can
get rid of $R_\infty$ and obtain
\begin{align}
\int^0_{-|t|_{\rm max}}\!\!\!\!\! dt \,
A(t) \bigg(\frac{e^2\kappa \kappa_g}{F_\pi^2 M_P^2}
-
\frac{e^2\kappa_{1}\kappa_3 t^2}{ M_P^4}\bigg)
&=\sum_{i}^{\rm s\mbox{-}ch} A_{s}(m_i,J_i) \,g_{\gamma^+ \gamma^+,i}\,g_{h^- h^-,i},
+
\sum_{i}^{\rm u\mbox{-}ch} A_{u}(m_i,J_i) \, g_{\gamma^+ h^-,i}^2,
\label{eq:smeared_rrhh}
\\
\hspace{-0.9cm} 
\int^{w_{\rm max}}_{0} \!\!\!\! dw \,
B(w) \bigg(
\frac{1}{M_P^2 w}+\frac{e^4\kappa_1^2 w}{M_P^2}+e^4c_{4\gamma}
\bigg)
&=\sum_{i}^{\rm s\mbox{-}ch}B_{s}(m_i,J_i) \,|g_{\gamma^+ \gamma^+,i}|^2
+
\sum_{i}^{\rm u\mbox{-}ch}B_{u}(m_i,J_i) \,|g_{\gamma^+ \gamma^-,i}|^2,
\label{eq:smeared_rrrr}
\\
\int^0_{-|t|_{\rm max}} \!\!\!\!dt \,
C(t) \frac{-1}{M_P^2 t}
&=\sum_{i}^{\rm s\mbox{-}ch}C_{s}(m_i,J_i) \,|g_{h^+ h^+,i}|^2
+
\sum_{i}^{\rm u\mbox{-}ch}C_{u}(m_i,J_i) \,|g_{h^+ h^-,i}|^2,
\label{eq:smeared_hhhh}
\\
\int^0_{-|t|_{\rm max}} \!\!\!\!dt \,
D(t) \frac{-1}{M_P^2 t}
&=\sum_{i}^{\rm s\mbox{-}ch}D_{s}(m_i,J_i) \,|g_{\gamma^+ h^-,i}|^2
+
\sum_{i}^{\rm u\mbox{-}ch}D_{u}(m_i,J_i) \,|g_{\gamma^+ h^+,i}|^2,
\label{eq:smeared_rhrh}
\end{align}
where the explicit forms of smeared Wigner $d$-functions $A_{s,u},B_{s,u},C_{s,u},$ and $D_{s,u}$ are presented in Appendix.~\ref{appendix:explicit_smeared}.

Toward deriving the  bounds, the first condition on the smeared Wigner $d$-functions is their positivity in terms of all allowed values of $m_i$ and $J_i$. This allows to 
derive the following simplified inequalities for the dispersion relations of the elastic processes:
\begin{align}
B_{s,u}(m_i,J_i)\geq 0
& \;\;\Rightarrow \!\!&
\int^{w_{\rm max}}_{0} \!\!\!\! dw \,
B(w) \bigg(
\frac{1}{M_P^2 w}+\frac{e^4\kappa_1^2 w}{M_P^2}+e^4c_{4\gamma}
\bigg)
&\,\geq\,\sum_{i}^{\rm s\mbox{-}ch}B_{s}(m_i,J_i) \,|g_{\gamma^+ \gamma^+,i}|^2,
\label{eq:simple_smeared_rrrr}
\\
C_{s,u}(m_i,J_i)\geq 0& \;\;\Rightarrow \!\!\!\!\!\!&
\int^0_{-|t|_{\rm max}} \!\!\!\!dt \,
C(t) \frac{-1}{M_P^2 t}
&\,\geq\,\sum_{i}^{\rm s\mbox{-}ch}C_{s}(m_i,J_i) \,|g_{h^+ h^+,i}|^2,
\label{eq:simple_smeared_hhhh}
\\
D_{s,u}(m_i,J_i)\geq 0& \;\;\Rightarrow \!\!\!\!\!\!&
\int^0_{-|t|_{\rm max}} \!\!\!\!dt \,
D(t) \frac{-1}{M_P^2 t}
&\,\geq\,\sum_{i}^{\rm s\mbox{-}ch}D_{s}(m_i,J_i) \,|g_{\gamma^+ h^-,i}|^2.
\label{eq:simple_smeared_rhrh}
\end{align}

\noindent
Secondly, in order to bound the inelastic process \eq{eq:smeared_rrhh},
we demand the conditions
\be
|A_s(m_i,J_i)|\leq \sqrt{B_s(m_i,J_i)C_s(m_i,J_i)}\ ,\ \ \ \ 
|A_u(m_i,J_i)|\leq D_s(m_i,J_i)\,,
\label{eq:conditions}
\ee
that respectively lead to
\begin{gather}
\sum_{i}^{\rm s\mbox{-}ch} |A_{s}(m_i,J_i) \,g_{\gamma^+ \gamma^+,i}\,g_{h^- h^-,i}|\leq 
\sum_{i}^{\rm s\mbox{-}ch} |\sqrt{B_{s}(m_i,J_i)} \, g_{\gamma^+ \gamma^+,i}|\,|\sqrt{C_{s}(m_i,J_i)}\, g_{h^- h^-,i}|,
\label{eq:A_leq_BC}
\\
\sum_{i}^{\rm u\mbox{-}ch} |A_{u}(m_i,J_i) \,g_{\gamma^+ h^-,i}^2|
\leq 
\sum_{i}^{\rm s\mbox{-}ch}D_{s}(m_i,J_i) \,|g_{\gamma^+ h^-,i}|^2
\,.
\label{eq:A_leq_D}
\end{gather}
Notice that the smeared Wigner $d$-functions $A_{s,u}$ are not required to be positive. In order to relate the RHS of \eq{eq:A_leq_BC}
to the RHS of \eq{eq:simple_smeared_rrrr} and \eq{eq:simple_smeared_hhhh},
we use the Cauchy--Schwarz inequality
\begin{align}
&\sum_{i}^{\rm s\mbox{-}ch}
|\sqrt{B_{s}(m_i,J_i)}\,g_{\gamma^+ \gamma^+,i}|
|\sqrt{C_{s}(m_i,J_i)}\,g_{h^- h^-,i}\big|
\nonumber
\\
&\leq
\sqrt{
\bigg(\sum_{i}^{\rm s\mbox{-}ch}
B_{s}(m_i,J_i)|g_{\gamma^+ \gamma^+,i}|^2
\bigg)
\bigg(
\sum_{i}^{\rm s\mbox{-}ch}
C_{s}(m_i,J_i)|g_{h^+ h^+,i}|^2
\bigg)}\,.
\end{align}
Collecting all the formulas discussed above, we can construct the following relation:
\begin{align}
&\bigg|\sum_{i}^{\rm s\mbox{-}ch} A_{s}(m_i,J_i) \,g_{\gamma^+ \gamma^+,i}\,g_{h^- h^-,i},
+
\sum_{i}^{\rm u\mbox{-}ch} A_{u}(m_i,J_i) \,g_{\gamma^+ h^-,i}^2
\bigg|
\nonumber
\\
&\leq
\sqrt{
\bigg(
\sum_{i}^{\rm s\mbox{-}ch}
B_{s}(m_i,J_i)|g_{\gamma^+ \gamma^+,i}|^2
\bigg)
\bigg(
\sum_{i}^{\rm s\mbox{-}ch}
C_{s}(m_i,J_i)|g_{h^+ h^+,i}|^2
\bigg)}+\sum_{J,\,i}^{\rm s\mbox{-}ch}D_{s}(m_i,J_i) \,|g_{\gamma^+ h^-,i}|^2,
\end{align}
leading to the inequality between the low-energy sides,
\begin{align}
&\bigg|\int^0_{-|t|_{\rm max}} \!\!\!\!dt \,
A(t) \bigg(\frac{e^2\kappa \kappa_g}{F_\pi^2 M_P^2}
-
\frac{e^2\kappa_{1}\kappa_3 t^2}{ M_P^4}\bigg)
\bigg|
\nonumber
\\
&\leq \; 
\sqrt{\bigg\{\int^{w_{\rm max}}_{0} \!\!\!\!dw \,
B(w)\, \bigg(
\frac{1}{M_P^2 w}+\frac{e^4\kappa_1^2 w}{M_P^2}+e^4c_{4\gamma}
\bigg)\bigg\}
\bigg\{\int^0_{-|t|_{\rm max}} \!\!\!\!dt \,
C(t) \frac{-1}{M_P^2 t}
\bigg\}
}
+
\int^0_{-|t|_{\rm max}} \!\!\!\!dt \,
D(t) \frac{-1}{M_P^2 t}\,,
\label{eq:EFT_inequ}
\end{align}
that provides a bound on the anomalous couplings.
This bound simplifies when neglecting subleading $1/M_P$ terms:
\begin{align}
\bigg|
\frac{\kappa \kappa_g }{F_\pi^2 M_P^2}
\int^0_{-|t|_{\rm max}} dt \,
A(t) 
\bigg|
\leq \; 
\sqrt{\frac{c_{4\gamma}}{M_P^2}
\int^{w_{\rm max}}_{0} dw \,
B(w)\, 
\int^0_{-|t|_{\rm max}} dt \,
C(t)\frac{-1}{t} }
\,.
\label{eq:EFT_inequ2}
\end{align}

The smearing functions must be chosen to satisfy all the conditions mentioned above. 
A choice of these functions is (see Appendix~\ref{appendix:positivity_of_smeared_d}  for a detailed discussion)
\begin{align}
A(t)&=C(t)=(1-\sqrt{-t/|t|_{\rm max}})^5,
\label{eq:smearing_A}
\\
B(w)&=4\,(1-\sqrt{w/w_{\rm max}})^5,
\label{eq:smearing_B}
\\
D(t)&=1.1\,J_0\Big(9\sqrt{-t/|t|_{\rm max}}\Big)\,(1-\sqrt{-t/|t|_{\rm max}})^5,
\label{eq:smearing_D}
\end{align}
where $J_0(x)$ is  the Bessel function  of the first kind.
We take  
$w_{\rm max}=|t|_{\rm max}/4$ to satisfy the inequality in Eq.~\eqref{eq:A_leq_BC}.\footnote{The factor 4 in the smearing function was introduced such that the integrals on the high-energy side yield the overall factor $|t|_{\rm max}$, rather than $|t|_{\rm max}/4$.} 
The smearing functions are chosen to satisfy all positivity conditions when $|t|_{\rm max}$ is smaller than the mass of the lightest heavy state appearing among the three elastic processes.
The value of $|t|_{\rm max}$ plays a crucial role, since the LHS of the bound \eq{eq:EFT_inequ2}  depends 
linearly on $\sqrt{|t|_{\rm max}}$, as can be easily deduced on dimensional grounds.  
Therefore the larger the value of $|t|_{\rm max}$, the  stronger the bound.
Nevertheless,  $|t|_{\rm max}$ 
cannot exceed  the mass of the lightest spin $J\geq 2$ UV state, denoted by $M_{J\geq 2}$, otherwise we lose  the positivity of the functions $B_{s,u}, C_{s,u}$ and $D_{s,u}$
(see Appendix~\ref{appendix:positivity_of_smeared_d}). For this reason, we  take 
\begin{align}
|t|_{\max}=M_{J\geq 2}^2\,.
\label{maxt}
\end{align}
We will later discuss a particular case where $M_{J\geq 2}$ corresponds to the mass of a $J=2$ state, which couples to $\gamma^+\gamma^+$ and $h^-h^-$, and can be exchanged in the processes $\gamma^+\gamma^+\rightarrow\gamma^+\gamma^+$ and $h^+h^+\rightarrow h^+h^+$.
This is consistent with causality bounds for higher-spin particles in the large $N_c$ limit~\cite{Kaplan:2019soo,Kaplan:2020ldi,Kaplan:2020tdz,Afkhami-Jeddi:2018apj}.

Plugging Eqs.~\eqref{eq:smearing_A}--\eqref{eq:smearing_D} and \eq{maxt} into Eq.~\eqref{eq:EFT_inequ2}, we get the bound
\begin{align}
M_{J\geq 2}\frac{|\kappa \kappa_g|}{F_\pi^2 M_P}
\lesssim 1.4 \sqrt{c_{4\gamma}}\ln^{\frac12}\bigg(\frac{M_{J\geq 2}}{M_{\rm IR}}\bigg),
\label{eq:bound_kkg}
\end{align}
where we have had to introduce the  IR cutoff $M_{\rm IR}$  to regularize the logarithmically divergent integral $\int dt\, C(t)/t$.  This requirement is a well-known feature of 4D gravity, also manifest in constraints from time delays in the eikonal limit~\cite{Camanho:2014apa} (and the time delay experienced by photons is positive not only at tree level but also at one-loop order).

\subsection{$\gamma^+\gamma^+\rightarrow h^- h^-$}
\label{sec:bound_rrhh2}

In addition to the bound in Eq.~\eqref{eq:bound_kkg}, we can derive an alternative bound on $\kappa\kappa_g$ from the process $\gamma^+\gamma^+ \rightarrow h^-h^-$,  which amplitude at low-energy,  written in terms of the variable $w$, is given by
\begin{align}
\mathcal{M}_{\gamma^+\gamma^+h^+h^+}= 
s
\bigg(
\frac{e^2\kappa \kappa_g \,s^2}{F_\pi^2 M_P^2\, s}
+\frac{e^2\kappa_1\,s^2}{M_P^2\,w}+\frac{e^2b_{2,0} \,s w}{ M_P^2}
+\cdots
\bigg),
\end{align}
where the first term arises from the $\eta$ exchange in the $s$-channel, the second one is for the $\gamma$ exchange in the $t$- and $u$-channel,
and $b_{2,0}$ is a  Wilson coefficient. 

Following the procedure outlined in  Sec.~\ref{sec:bound_rrhh1}, 
we get the bound
\begin{align}
M_{J\geq 2}\,
\bigg|\frac{\kappa \kappa_g}{F_\pi^2 M_P}+
\frac{b_{2,0} M^2_{J\geq 2}}{8M_P}\bigg|
\lesssim 1.4 \sqrt{c_{4\gamma}}
\ln^{\frac12}\bigg(\frac{M_{J\geq 2}}{M_{\rm IR}}\bigg),
\end{align}
that is similar to Eq.~\eqref{eq:bound_kkg} but  contains the additional contribution from $b_{2,0}$.

\subsection{$\gamma^+ h^+ \rightarrow \gamma^- h^+$}
\label{sec:bound_rhrh1}

\begin{figure}[t!]
\centering
\includegraphics[scale=1.1]{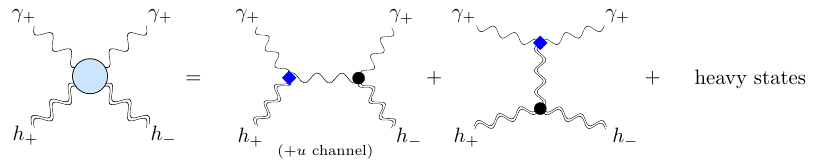}
\caption{Contributions to the process $\gamma^+ h^+\rightarrow \gamma^-h^+$. 
}
\label{fig:4pt_kappa1}
\end{figure}

Having established the bounds on $\kappa\kappa_g$, we also derive a bound on the non-minimal coupling $\kappa_1$. For this purpose we use the process $\gamma^+ h^+ \rightarrow \gamma^- h^+$
whose low-energy amplitude is given by
\begin{align}
&\mathcal{M}_{\gamma^+ h^+ \gamma^+ h^-}=
s^{2} u^{2} t\,
\bigg(
\frac{e^2\kappa_{1}}{M_P^2 s t u}
+\cdots
\bigg),
\end{align}
where $\kappa_1$ appears due to the $\gamma$ exchange in     the $s$- and $u$-channels,
and $h$ exchange in the  $t$-channel  --see Fig.~\ref{fig:4pt_kappa1}. 
The $s\leftrightarrow u$ symmetric property of the amplitude leads  to the following simple form of the dispersion relation with $k=2$ and $l=1$ or $2$:
\begin{align}
&\frac{e^2\kappa_{1}}{M_P^2}
=
\Bigg\langle \frac{(2m^2+t)d^J_{3,1}(1+2t/m^2)}{m^2
(m^2+t)^2}\Bigg\rangle^{g_{\gamma^+ h^+}}_{g_{\gamma^+ h^-}}+R_\infty (t)\,.
\label{eq:dispersion_rhrh2}
\end{align}
To  bound  $\kappa_1$, we need to compare this 
to the elastic $\gamma^+h^-\rightarrow \gamma^+ h^-$, whose contributions are shown at the bottom of Fig.~\ref{fig:4pt_for_bounds} and its dispersion relation is given in Eq.~\eqref{eq:dispersion_rhrh}.

Next, we integrate over the dispersion relations Eqs.~\eqref{eq:dispersion_rhrh2} and~\eqref{eq:dispersion_rhrh} with  some arbitrary smearing functions, $E(t)$ and $F(t)$, respectively:
\begin{align}
\int^0_{-|t|_{\rm max}} dt \,
E(t) \frac{e^2\kappa_{1}}{M_P^2}
&=\sum_{i}^{\rm s\mbox{-}ch}E_s(m_i,J_i) \,g_{\gamma^+ h^+,i}\,g_{\gamma^+ h^-,i},
\label{eq:smeared_rhrh2}
\\
\int^0_{-|t|_{\rm max}} dt \,
F(t) \frac{-1}{M_P^2 t}
&=\sum_{i}^{\rm s\mbox{-}ch}F_{s}(m_i,J_i) \,|g_{\gamma^+ h^-,i}|^2
+
\sum_{i}^{\rm u\mbox{-}ch}F_{u}(m_i,J_i) \,|g_{\gamma^+ h^+,i}|^2,
\label{eq:smeared_rhrh3}
\end{align}
where  in Eq.~\eqref{eq:smeared_rhrh2}
we have combined  the $s$- and $u$-channels using the $s\leftrightarrow u$ symmetry.
The explicit forms of $E_s$ and $F_{s,u}$ are presented in Appendix~\ref{appendix:explicit_smeared}. As in the procedure for bounding $\kappa\kappa_g$ in Sec.~\ref{sec:bound_rrhh1}, the smeared Wigner $d$-function $E_s$ for the  process $\gamma^+h^+\rightarrow \gamma^-h^+$ can be either positive or negative depending on $i$. With this fact in mind, we impose the following conditions on $E_s$ and $F_{s,u}$:
\be
|E_s(m_i,J_i)|\leq \sqrt{F_{s}(m_i,J_i)F_{u}(m_i,J_i)}
\quad \mbox{and} \quad F_{s}(m_i,J_i),F_{u}(m_i,J_i) \geq 0\,,
\label{eq:E_F_condition}
\ee
that lead to the  inequality
\begin{align}
 \sum_{i}^{\rm s\mbox{-}ch}|E_s(m_i,J_i) \,g_{\gamma^+ h^+,i}\,g_{\gamma^+ h^-,i}|
\leq 
\sum_{i}^{\rm s\mbox{-}ch} |\sqrt{F_{s}(m_i,J_i)} \, g_{\gamma^+ h^-,i}|\,|\sqrt{F_{u}(m_i,J_i)}\, g_{\gamma^+ h^+,i}|\,.
\label{eq:EF_condition_and_inequality}
\end{align}
This inequality remains valid even without the absolute value bars on the LHS. Due to angular momentum conservation, the summation on the RHS must start from $J=3$, as indicated by the $s$-ch symbol. Using the Cauchy--Schwarz inequality,
\begin{align}
&
\sum_{i}^{\rm s\mbox{-}ch}
|\sqrt{F_{s}(m_i,J_i)}\,g_{\gamma^+ h^-,i}|
|\sqrt{F_{u}(m_i,J_i)}\,g_{\gamma^+ h^+,i}\big|
\nonumber
\\
&\leq
\sqrt{
\bigg(\sum_{i}^{\rm s\mbox{-}ch}
F_{s}(m_i,J_i)|g_{\gamma^+ h^-,i}|^2
\bigg)
\bigg(
\sum_{i}^{\rm u\mbox{-}ch}
F_{u}(m_i,J_i)|g_{\gamma^+ h^+,i}|^2
\bigg)},
\label{eq:cauchy2}
\end{align}
in \eq{eq:EF_condition_and_inequality}, we get
\begin{align}
\bigg|\sum_{i}^{\rm s\mbox{-}ch}E_s(m_i,J_i) \,g_{\gamma^+ h^+,i}\,g_{\gamma^+ h^-,i}
\bigg|
\leq
\sqrt{
\frac12
\bigg(
\sum_{i}^{\rm s\mbox{-}ch}F_{s}(m_i,J_i) \,|g_{\gamma^+ h^-,i}|^2
+\sum_{i}^{\rm u\mbox{-}ch}F_{u}(m_i,J_i) \,|g_{\gamma^+ h^+,i}|^2
\bigg)^2}\,.
\end{align}
This, together with Eqs.~\eqref{eq:smeared_rhrh2} and~\eqref{eq:smeared_rhrh3}, leads to
\begin{align}
\bigg|
\frac{e^2\kappa_{1}}{M_P^2}
\int^0_{-|t|_{\rm max}} dt \,
E(t) 
\bigg|
\leq 
\frac{1}{\sqrt{2}M_P^2}\int^0_{-|t|_{\rm max}} dt \,
F(t) \frac{-1}{t},
\label{eq:rare_EF_bound}
\end{align}
implying the linear $|t|_{\rm max}$ dependence of the bound.

One choice of the smearing functions satisfying the conditions in Eq.~\eqref{eq:E_F_condition} is given by\footnote{For the function $E(t)$ the overall factor $(-t/|t|_{\max})^{5/8}$ was  chosen
to satisfy the condition Eq.~\eqref{eq:EF_condition_and_inequality}
at large $J$ (see Appendix~\ref{appendix:high_J_asym}).}
\begin{align}
E(t)&=(-t/|t|_{\rm max})^{5/8}(1-\sqrt{-t/|t|_{\rm max}})^5,
\label{eq:smearing_E}
\\
F(t)&=J_0(9\sqrt{-t/|t|_{\rm max}})(1-\sqrt{-t/|t|_{\rm max}})^5\,.
\label{eq:smearing_F}
\end{align}
As discussed in Sec.~\ref{sec:bound_rrhh1}, we must take the largest possible value for 
$|t|_{\rm max}$ without violating the positivity of $F_{s,u}$. This corresponds to  (see Appendix~\ref{appendix:positivity_of_smeared_d} for details)
\begin{align}
|t|_{\max}=M_{J\geq 3}^2,
\label{eq:t_M3}
\end{align}
which is the mass of the lightest spin $J\geq 3$ UV states. 
In the particular case where this UV state has
$J=3$, and couples to $\gamma^+h^+$ and $\gamma^-h^+$, 
this is the first massive UV state encountered in the processes $\gamma^+h^+\rightarrow \gamma^+h^+$ and $\gamma^-h^+\rightarrow \gamma^-h^+$. 

Plugging Eqs.~\eqref{eq:smearing_E} and \eqref{eq:smearing_F} into Eq.~\eqref{eq:rare_EF_bound} with \eqref{eq:t_M3} we obtain the bound
\begin{align}
e^2|\kappa_{1}| M^2_{J\geq 3}
\lesssim 226\ln \bigg(\frac{M_{J \geq 3}}{M_{\rm IR}}\bigg),
\label{eq:bound_k1}
\end{align}
where the IR cutoff $M_{\rm IR}$ is required to regulate the  logarithmic divergence in $\int dt F(t)/t$.

\subsection{$\gamma^+ h^- \rightarrow \gamma^+ h^-$}
\label{sec:bound_rhrh2}

The elastic scattering $\gamma^+h^- \rightarrow \gamma^+h^-$, whose EFT amplitude is given in Eq.~\eqref{eq:EFT_rhrh}, can solely yield the same type of bound as Eq.~\eqref{eq:bound_k1}. 
As shown in Table~\ref{tab:anomaly_contents}, the appearance of $\kappa_1^2$ in the dispersion relation at $k=3$ allows us to bound it by comparing it with the dispersion relation at $k=2$. This procedure is the same as that used to bound $\kappa_g$ from the elastic scattering $h^+h^+\rightarrow h^+h^+$ in Ref.~\cite{Dong:2024omo}. We leave the explicit computation to the reader, as the full procedure has already been presented in Ref.~\cite{Dong:2024omo}.

\section{Phenomenological implications}
\label{sec:pheno}

There is a large class of models whose spectrum   contains  a pseudoscalar, a photon and a graviton with the couplings  of \eq{eq:3pt_anomaly}.
The most popular examples are axion models, where the axion is a pseudo-Goldstone boson arising from the spontaneous breaking of a $U(1)_{\rm PQ}$ symmetry, which can be identified with $\eta$ and generically couples to photons and gravitons as in \eq{eq:3pt_anomaly}.
 They can either be generated from UV physics or from a  mixing of the axion with the QCD $\pi^0$ and $\eta'$.

In the bound Eq.~\eqref{eq:bound_kkg}, $M_{J\geq 2}$ can  be  interpreted as  
 the scale at which states with $J\geq 2$ must appear in the theory coupled linearly to $\gamma\gamma$ and $hh$
--see \eq{maxt}--   and therefore corresponds to 
an  upper bound for the cutoff scale of  ${\cal L}_{\rm EFT}(\eta,\gamma,h)$.
We will refer to this cutoff scale as  $\Lambda_{\text{caus}}$.
From  \eq{eq:bound_kkg}, taking $M_{\rm IR}= O(M_{J\geq 2})$,  the scale  $\Lambda_{\text{caus}}$ is    parametrically given by
\be 
 \Lambda_{\text{caus}}\sim M_P \frac{\sqrt{c_{4\gamma}}\, F_\pi^2}{ |\kappa \kappa_g|}\,.
\label{cutoff}
\ee
This   cutoff scale  can be smaller than $\Lambda_{\text{pert}}$, 
the naive perturbative cutoff scale
of ${\cal L}_{\rm EFT}(\eta,\gamma, h)$  defined as the scale at which loops become of tree-level order.
$\Lambda_{\text{pert}}$ can be estimated by the energy scale at which  the contribution mediated by $\eta$  to the amplitude of $\gamma\gamma\to hh $ (first term of \eq{eq:EFT_rrhh}) becomes $O(1)$.
We find
\be 
\Lambda_{\text{pert}}\sim   \sqrt{\frac{M_PF_\pi}{e\, (\kappa  \kappa_g)^{1/2} }} \,.
\label{pertbound}
\ee
For sufficiently small $c_{4\gamma}$ or $e$,   one  finds $\Lambda_{\text{caus}}$  smaller than $\Lambda_{\text{pert}}$.

Since $c_{4\gamma}$ is expected to be generated by physics at the cutoff,  we can assume
$c_{4\gamma}\sim 1/\Lambda^4_{\rm caus}$. In this case  \eq{cutoff} leads to
\be 
 \Lambda_{\text{caus}}\sim \left( \frac{M_P F_\pi^2}{ |\kappa \kappa_g|}\right)^{1/3}\,,
\label{cutoffalt}
\ee
that tells us that for $\kappa\kappa_g\sim 1$ the cutoff scale lies between $M_P$ and $F_\pi$. 

 We can also  interpret \eq{eq:bound_kkg} as a bound on the smallest value of $\gamma\gamma\to\gamma\gamma$ at low-energies.
Taking as an example the largest value for the causal cutoff, $\Lambda_{\text{caus}}\sim M_P$, one finds
\be
c_{4\gamma}\gtrsim \left(\frac{ \kappa \kappa_g}{ F_\pi^2}\right)^2\,.
\ee

\subsection{Strongly-coupled  gauge theories}
\label{largeNgauge}

Strongly-coupled gauge theories with $N_F$ fermions  and   spontaneous breaking of   global symmetries, e.g., 
$U(1) \times U(1)\to U(1)$,  contain  a light pseudoscalar state, $\eta$, in their spectrum.\footnote{In theories like QCD this is the chiral symmetry  breaking
$U(1)_L \times U(1)_R\to U(1)_V$.
In this case  the axial $U(1)_A$ current $J^\mu_A = \sum_f\bar{f} \gamma^\mu \gamma^5 f$ is not conserved at the quantum level due to the axial anomaly,  and  $\eta$ gets a mass
of order $m^2_\eta\sim N_F M^4/F_\pi^2$ \cite{Witten:1979vv, Veneziano:1979ec}.
Nevertheless, in  the large-$N_c$ limit  
this mass is  suppressed  since $F^2_\pi\sim N_c M^2\to \infty$ and 
then  $m^2_\eta\to 0$.}

We can estimate the size of the parameters as a function of $N_F$ and $N_c$, the number of fermions and  ``colors" respectively, of the gauge group.
In the large-$N_c$ limit the scaling of couplings follows the planar expansion~\cite{tHooft:1973alw,Witten:1979kh},
and  bootstrap bounds on strongly-coupled  theories in this limit   have been derived recently
\cite{Albert:2022oes,Fernandez:2022kzi,Albert:2023jtd,Ma:2023vgc,Li:2023qzs,Albert:2023seb} 
(see also~\cite{
Afkhami-Jeddi:2018apj,Kaplan:2019soo,Kaplan:2020ldi,Kaplan:2020tdz}).
For the particular case  of $SU(N_c)$ gauge theories,  with fermions in the fundamental representation,  we have 
\be
\kappa\sim \kappa_g \sim N_c\sqrt{N_F}\ , \ \  
F_\pi \sim \sqrt{N_c}\ , \ \
c_{4\gamma} \sim N_c N_F
\,.
\label{anomaly}
 \ee
As stated in \cite{Dong:2024omo}, demanding that loop corrections to the graviton propagator are smaller than tree-level, leads to  a  bound on  the theory  around ${M_P}/{N_c}\equiv \Lambda_{\rm QG}$.
Plugging the relations~\eqref{anomaly} with $M_P=\Lambda_{\rm QG}N_C$ into \eq{cutoff}, we get
\be
 \Lambda_{\text{caus}}\sim \Lambda_{\text{QG}}  \sqrt{\frac{N_c}{N_F}}\,.
\label{boundncnf}
\ee
We see that $\Lambda_{\text{caus}}$ is larger or of order of   
 $\Lambda_{\text{QG}}$ and therefore no meaningful  bound can be derived for the $J\geq2$  states.

Our analysis can  be used however to put bounds on 3-pt couplings generated from these models. For example, \eq{eq:bound_k1} leads  to a bound
on $\kappa_1$ as a function of the  mass of the lightest $J\geq 3$ state that parametrically is given by
 \be
e^2|\kappa_1|\lesssim  \frac{1}{M^2_{J\geq 3}}
\,.
\label{boundk1}
\ee
Since $\kappa_1$ can be generated from  loops of charged  fermion, it scales as $\sum_F q_F^2 N_c$, where $q_F$ is the charge of the fermion, and therefore the bound
\eq{boundk1} seems very  restrictive in the large-$N_c$ limit.
Nevertheless, one must take into account that $e^2$ is also sensitive to $\sum_F q_F^2  N_c$ as  fermions  affect the photon
propagator. In particular, one has 
\be
\frac{1}{e^2(\mu)}=\frac{1}{e^2(\mu_0)}+\sum_F \frac{q_F^2 N_c}{6\pi^2} \ln\mu_0/\mu\,.
\label{running}
\ee 
From \eq{running}, we can estimate a lower limit for the gauge coupling at $\mu\lesssim \mu_0$,
$e^2\gtrsim  1/\sum_F q_F^2 N_c$, 
 that 
used in \eq{boundk1} leads to
 \be
\frac{|\kappa_1|}{\sum_F q_F^2  N_c }\lesssim  \frac{1}{M^2_{J\geq 3}}\,.
\label{boundk2}
\ee
The above bound is specially interesting for strongly-coupled models with  an holographic description (see later for details).
In these models the   $J\geq 3$ states can be parametrically much heavier than  the $J<3$ states,  and therefore \eq{boundk2}
tells us that $\kappa_1$ must be suppressed.

\subsection{5D models and holography}

Another  class of  models whose low-energy  spectrum consists  of $\eta$, $\gamma$, and $h$ are 
5D models   with 
$U(1)_V\times U(1)_A$ gauge invariance. Its Lagrangian can be written as
\bea
{\cal L}_5&=&\int d^4x\, dz\,  \sqrt{-g}  \bigg[M_{P5}^3 {\cal R}+M_5 \bigg(V_{MN}^2+A_{MN}^2+c_{4V}  V_{MN}^4+\cdots\bigg)\nonumber\\
&&\hskip2.5cm
  +\,  \kappa_5\, \epsilon_{MNPQR} A^M F^{NP} F^{QR}
+\kappa_{g5}\, \epsilon_{MNPQR} A^M {\cal R}^{NPA}_B {\cal R}^{QRB}_A\bigg] \,,
\label{lagrangian5}
\eea
where $M,N,...=\mu,4$;  $V_{MN}$ and $A_{MN}$ are 
the field-strength tensors of  the $U(1)_V$ and $U(1)_A$, while ${\cal R}$ and ${\cal R}^{MNPQ}$
are respectively the   Ricci scalar and Riemann tensor;
$M_{P5}$  and $1/M_{5}$ are respectively the 5D Planck scale and 
 gauge-coupling squared, and $\kappa_5$, $\kappa_{g5}$ are the dimensionless coefficient of  the Chern-Simons 
terms related to  $U(1)_A$ anomalies.
The Lagrangian \eq{lagrangian5} is derived as an expansion in fields and derivatives, 
with Dim$[A_M]$=Dim$[V_M]=1$, and we have only included
one term of dimension $8$, $V_{MN}^4$,
that will play an important role in our analysis.

We will assume that the extra dimension  $z$ is compactified in a segment (orbifold) and will  take as  boundary conditions  $\partial_z V_\mu=0$ and $A_\mu=0$ at  both boundaries. For the graviton the boundary condition depends on the metric  and it is usually referred to as the Israel junction condition~\cite{Israel:1966rt}.
At  low-energies,
$E\ll 1/L$, the model has the following massless states:  $V_\mu$ that we identify with the photon,  the fifth-component of $A_M$, $A_5$ (a pseudoscalar) that we identify with $\eta$, 
and  the  graviton $h_{\mu\nu}$. 
The theory also contains massive states, the Kaluza-Klein (KK) modes, 
for $V_\mu$, $A_\mu$, and $h_{\mu\nu}$. 
The Chern-Simons terms proportional to $\kappa_5$ and  $\kappa_{g5}$ leads respectively to  the couplings of  \eq{eq:3pt_anomaly}.
 
 Let us start considering the case in which   the 5D space is flat.
We  then have  \cite{Barbieri:2003pr}
\be
F^2_\pi\sim M_5/L\ , \ \  M_P^2 \sim M^3_{P5}L\,.
\label{flatcase}
\ee
In this theory $c_{4\gamma}$ could either be generated from integrating the KK  states or from the higher-dimensional operator $V^4_{MN}$. 
We find that in flat space there is no coupling of a single  KK mode to  two massless $V_\mu$ (the photon), due to
translational invariance in the 5th dimension. Therefore, $c_{4\gamma}$ cannot be generated at tree-level
from integrating out the KK modes, and  its only contribution  arises from $c_{4V}$, i.e.  $c_{4\gamma}\propto c_{4V}$. 
Given this  and \eq{flatcase}, we have that   the causal bound \eq{cutoff} can be written as 
\be
\Lambda_{\rm caus}\sim \frac{\sqrt{ M_5^{3} M^{3}_{P5}c_{4V}}}{\kappa_5 \kappa_{g5}}\,.
\label{flat}
\ee
The scale $\Lambda_{\rm caus}$ corresponds to the  mass of the lightest $J\geq 2$ state that couples to $\gamma \gamma$, that could in principle   be the lightest KK graviton, of mass $\sim 1/L$. 
Nevertheless, as we mentioned before, KK gravitons do not  couple  to $\gamma\gamma$ and 
therefore $\Lambda_{\rm caus}$ must be associated to the mass of  a new state.
Additionally,  since $c_{4V}$  is expected to  scale  with   the 5D cutoff scale of the theory,  which  can be identified  with $\Lambda_{\rm caus}$, i.e.  $c_{4V}\propto 1/\Lambda_{\rm caus}^4$, we   obtain from \eq{flat}
\be
\Lambda_{\rm caus}\sim \frac{\sqrt{ M_5 M_{P5}}}{(\kappa_5 \kappa_{g5})^{1/3}}\,.
\label{flat2}
\ee
Notice that this bound  does not depend on $L$, so it applies even in the decompactification limit $L\to\infty$.

\begin{figure}[t!]
\centering
\hspace{-1cm}
\includegraphics[scale=0.75]{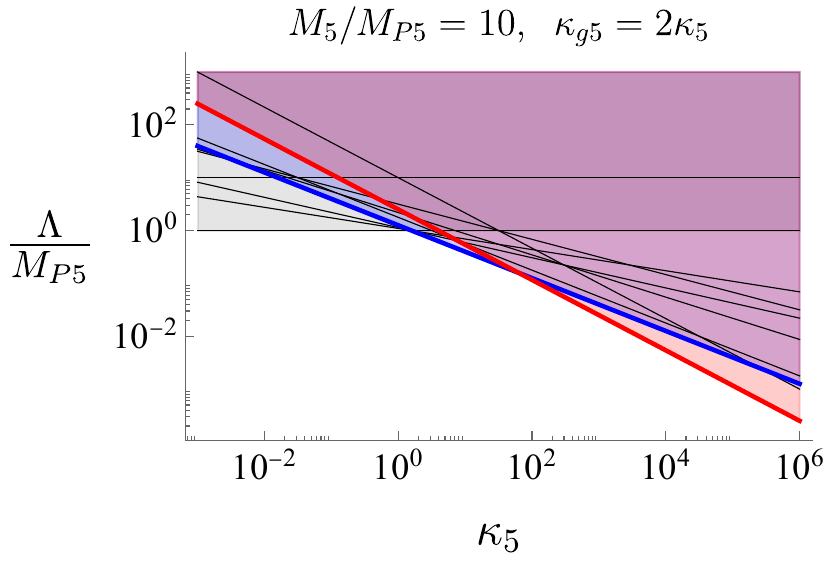}
\caption{Bounds on the cutoff scale of the 5D model \eq{lagrangian5}. The black lines corresponds to the perturbative bounds \eq{pertcut}, while the red (blue) corresponds to the bound from causality \eq{flat2} (\eq{flat3}).}
\label{fig:bounds}
\end{figure}

How  \eq{flat2} compares with other bounds?
We can naively get bounds on the cutoff scale of the 5D Lagrangian \eq{lagrangian5} 
 by demanding that loops are smaller than tree-level effects, i.e.~naive dimensional analysis~\cite{Manohar:1983md}.
 In particular, we can estimate 
 the cutoff scale $\Lambda_{5\, \rm pert}$ as the scale at which loops in   3-vertices involving $V_M$ and $h_{MN}$  
become  of order of their tree-level value. This leads to 
\be
\Lambda_{5\,\rm pert }={\rm Min} 
\Bigg\{M_5,M_{P5},
\frac{M_{5}}{\kappa_5^{2/3}},
\frac{\sqrt{M_5M_{P5}}}{\kappa_5^{1/3}},
\!\!\!\!\!\!{\phantom{\frac{M}{M}}}^{10}\!\!\!\!\sqrt{\frac{M_5 M_{P5}^9}{\kappa_{g5}^2}},
\!\!\!\!\!\!{\phantom{\frac{M}{M}}}^5\!\!\!\!\sqrt{\frac{M_5^2 M_{P5}^3}{\kappa_5 \kappa_{g5}}},
\!\!\!\!\!\!{\phantom{\frac{M}{M}}}^7\!\!\!\!\sqrt{\frac{M_5 M_{P5}^6}{\kappa_{g5}^2}}
\Bigg\}\,.
\label{pertcut}
\ee
Depending on the parameters of the model, we find regions where the causal cutoff \eq{flat2} is smaller (and therefore more relevant) than the perturbative cutoffs \eq{pertcut}.
This can be appreciated in Figure~\ref{fig:bounds}, where for  completeness, we have also included the 
 causal bound obtained in \cite{Dong:2024omo}:
\be
\Lambda_{\rm caus}\sim  
\!\!\!\!\!\!{\phantom{\frac{M}{M}}}^4\!\!\!\!\sqrt{\frac{M_5 M_{P5}^3}{\kappa_{g5}^2}}
\,.
\label{flat3}
\ee
It is easy to see for example that the causal upper bound \eq{flat2} becomes the smallest   of all
in the limit  $\kappa_g\sim \kappa\to \infty$ and $M_{5}/M_{P5}\sim \kappa^d$  for $0<d< 2/3$.

Let us now consider the case of warped extra dimensions,
for example, the case of an  AdS$_5$ metric, $ds^2=(L/z)^2 (\eta_{\mu\nu}dx^\mu dx^\nu-dz^2)$, being $L$ now  the AdS curvature radius.
We also consider that the  extra dimension  is limited  by  two branes, one at $z =1/\Lambda_{\rm UV}$  (UV-boundary) and another at $z =1/\Lambda_{\rm IR}$  (IR-boundary)
with $\Lambda_{\rm UV}\gg \Lambda_{\rm IR}$ \cite{Randall:1999ee}. In holographic language, this setup is dual to a large-$N_c$ CFT with IR breaking~\cite{Aharony:1999ti}, which also induces corrections to the graviton propagator~\cite{Arkani-Hamed:2000ijo}.
The KK masses are of order 
$\sim \Lambda_{\rm IR}$ and 
 the scale at which gravity loops are important is given by $\Lambda_{\rm QG}\sim \Lambda_{\rm UV}$. We now have $F^2_\pi\sim (M_5L) \Lambda_{\rm IR}^2$ and $M^2_P\sim (M_{P5} L)^3 \Lambda_{\rm UV}^2$. Contrary to the flat case, now $c_{4\gamma}$ can be generated from KK modes. For abelian theories, there is no 3 gauge vertices and the only contribution to $c_{4\gamma}$  comes from the exchange of KK gravitons. One 
gets the estimate 
\be
c_{4\gamma}\sim \frac{M_{5}^2}{M_{P5}^3L}\frac{1}{\Lambda_{\rm IR}^4}\,. 
\ee
Furthermore,   now $\Lambda_{\rm caus}$ can  be  associated with  the mass of the lightest KK graviton  $\sim \Lambda_{\rm IR}$.
Taking all these aspects into account,   we derive  from \eq{cutoff} the bound
\be
\frac{\kappa_5 \kappa_{g5}}{ (M_5L)^2}\lesssim \frac{\Lambda_{\rm UV}}{\Lambda_{\rm IR}}\,,
\label{wb}
\ee
which is naturally interpreted in the dual 4D theory via large-$N$ scaling~\cite{Aharony:1999ti}.
Finally, for explicit D3/D7-type holographic realizations, see ~\cite{Kruczenski:2003be} and \cite{Afkhami-Jeddi:2018own} for a conformal-collider perspective in holographic CFTs.

The AdS/CFT correspondence relates theories in $\text{AdS}_5$ to 4D strongly-coupled gauge theories in the limit $N_c \to \infty$ and $\lambda \equiv g^2_{\rm YM} N_c \to \infty$. In these limits, the $J>2$ states (associated with string excitations) become infinitely heavy, and the light spectrum consists only of states with $J=0,1,2$, as described by \eq{lagrangian5}.
The AdS/CFT correspondence dictates
\be
M_5 L \sim N^r_c, \quad
\kappa_5, \kappa_{g5} \sim N^r_c, \quad
\text{and thus} \quad \frac{\kappa_5 \kappa_{g5}}{(M_5 L)^2} \sim N_c^0\,,
\ee
where $r=1$ for fermions in the fundamental representation and $r=2$
for fermions in representations whose number of states grows as $N_c^2$, such as the symmetric or antisymmetric representations.  In both cases, we find that  \eq{wb}  provides an $N_c$-independent bound on the parameters of the model.

The bound \eq{boundk1} is also particularly interesting for strongly-coupled models in the limit $\lambda \to \infty$. In this regime, in the holographic dual, states with $J \geq 3$ become infinitely heavy,  and therefore \eq{boundk1} implies that $\kappa_1$ must vanish.

\subsection{Astrophysical constraints on the non-minimal coupling $\kappa_1$}
\label{sec:astro-constraints}

The $h\gamma\gamma$ non-minimal coupling $\kappa_1$ in \eq{eq:CPeven_Lagrangian} modifies the photon propagation in curved spacetime, leading to birefringence and helicity-dependent Shapiro delays.
The most stringent bound on $\kappa_1$ is obtained from the timing of binary pulsar PSR B1534+12 signals that gives $e^2|\kappa_1|\lesssim6\,\mathrm{km}^2$~\cite{Stairs:2002cw,Stairs:1999zr,Prasanna:2003ix}, and complementary constraints based on other astrophysical observations can be found in Ref.~\cite{Wu:2017yjl,Jana:2021lqe,Carballo-Rubio:2025zwz,Chen:2021lvo}.

Astrophysical bounds on $\kappa_1$ are typically of order
$e^2|\kappa_1|/r_g^2 \lesssim O(1)$, meaning that deviations must remain below the curvature scale set by the gravitational radius $r_g$, which for black holes corresponds to the horizon radius.
For M87* ($r_g\!\sim\!10^{9}\,$km), this implies $e^2|\kappa_1|\!\lesssim\!10^{18}$\,km$^2$, while for stellar-mass black holes ($r_g\!\sim\!10\,$km) one finds $e^2|\kappa_1|\!\lesssim\!10^2$\,km$^2$.

A $e^2|\kappa_1|\sim 6\,\mathrm{km}^2$, implies, using our bound \eq{eq:bound_k1}, 
that there should be new physics  at energies
\be
\Lambda\sim\!8\times10^{-11}\,\mathrm{eV}\,.
\ee
This could correspond to new particles with $J \geq 3$ coupled to $\gamma h$, or, in the case where new physics appears at the loop level, to new states coupled to $\gamma$.
Such low-mass states should therefore have been observed in other experiments, for example, through their effects on stellar evolution.
Our bounds indicate that searches for modifications of GR in binary pulsar timing or similar experiments are not competitive with existing experimental constraints.

\section{Conclusion}
\label{sec:conclusion}

We have obtained causality bounds by bootstrapping all possible $2\to2$ amplitudes EFTs consisting of a pseudoscalar $\eta$, a photon $\gamma$, and a graviton $h$, and whose interactions are dictated by $U(1)_\eta$ anomalies, as shown in Eq.~\eqref{eq:3pt_anomaly}.

To bootstrap amplitudes containing the gravitational $t$-pole divergence, we integrated the dispersion relation over $t$ or $w = tu/s$ using a weighting function that smears out such divergent contributions.
This allowed us to derive, from the process $\gamma^+\gamma^+ \rightarrow h^+h^+$, the bound Eq.~\eqref{eq:bound_kkg} on $\kappa\kappa_g$ as a function of the mass of the $J\geq2$ states in the EFT’s UV completion.
For EFTs with anomalies, such as axion models, this bound can be interpreted as a causal UV cutoff of ${\cal L}_{\rm EFT}(\eta,\gamma,h)$ [see Eq.~\eqref{cutoff}].
We have shown that this scale can lie below the perturbative UV cutoff estimated in Eq.~\eqref{pertbound}.

In addition, from the process $\gamma^+h^+ \rightarrow \gamma^-h^+$, we derived a bound on the non-minimal $h\gamma\gamma$ coupling $\kappa_1$, given in Eq.~\eqref{eq:bound_k1}, as a function of the mass of the lightest $J\geq3$ UV state.

We further analyzed the implications of our results in strongly-coupled gauge theories and in 5D models with Chern–Simons terms.
In the latter case, we showed that the causal cutoff can be smaller than the naive perturbative cutoff scale $\Lambda_{\rm 5 pert}$ of Eq.~\eqref{pertcut}, especially in the limit $\kappa_g\sim\kappa\to\infty$ and $M_5/M_{P5}\sim\kappa^d$ for $0<d<2/3$.
We also discussed the implications of these bounds for warped extra-dimensional models, which, thanks to the AdS/CFT correspondence, are related to strongly-coupled gauge theories in the large-$N_c$ and large-$\lambda$ limits.
Our bound Eq.~\eqref{eq:bound_kkg} provides an $N_c$-independent constraint on the anomaly coefficients [\eq{wb}], while Eq.~\eqref{boundk1} implies that $\kappa_1\to0$ as $\lambda\to\infty$.

Finally, we have shown that modifications of General Relativity arising from the photon–graviton coupling $\kappa_1$ would require new physics at very low energies, $\sim10^{-10}\text{eV}$, in order to be detectable in current or near-future binary pulsar timing experiments.

Although our bounds were derived in 4D, 
it would be interesting to obtain these results directly in 5D where log divergences are absent, and analyze their implications for 5D holographic duals of hydrodynamical systems at finite temperature and chemical potential~\cite{Cremonini:2009sy,Delsate:2011qp,Das:2005za,Donos:2012wi,Cai:2012mg,Megias:2013joa,Azeyanagi:2015gqa,Bhattacharyya:2016knk,Liu:2016hqb,Baggioli:2024zfq}.
For example, it would be interesting to explore whether violations of our causal bounds are connected to instabilities  in these theories.
One could also attempt to extend our analysis to theories involving scalar fields, as in~\cite{Serra:2022pzl}.
We leave these directions for future work.

\section*{Acknowledgments}
We thank Yifan Chen and Francesco Sciotti for discussions.
AP and ZD have been partly
supported by the research grants 2021-SGR-00649 and PID2023-146686NB-C31. JJ was supported by a KIAS Individual Grant (QP090001) through the Quantum Universe Center at the Korea Institute for Advanced Study.

\appendix

\section{Improved dispersion relation for $t \leftrightarrow u$ symmetric amplitudes}
\label{appendix:dispersion_relation_at_fixed_w}

The  $k = 2$ dispersion relation for the amplitude $\gamma^+\gamma^+\to \gamma^+\gamma^+$
at fixed $t$  
involves infinitely many low-energy Wilson coefficients from contact terms.
This can be avoided by keeping  another variable fixed instead of $t$ in the dispersion relations, as shown 
in \cite{Sinha:2020win,Li:2023qzs,Berman:2024kdh,Pasiecznik:2025eqc,EliasMiro:2025rqo} for $s$-$t$-$u$ symmetric amplitudes.
We present here a similar approach, valid for $t \leftrightarrow u$ symmetric amplitudes, 
which was {\it in parallel}  elaborated in \cite{Bellazzini:2025shd}.

We consider the particular case where
\be
w \equiv \frac{tu}{s},
\ee
is taken fixed in a $t \leftrightarrow u$ symmetric amplitude.
The expressions for $t(s, w)$  and $u(s, w)$   are given in this case by
\be
t=-s
\bigg(
\frac{1-\sqrt{1-4w /s}}{2}
\bigg)
\,,\quad 
u=-s
\bigg(\frac{1+\sqrt{1-4w/s}}{2}
\bigg)\,,
\label{eq:t_and_u}
\ee
where $w>0$ in the physical region.\footnote{The signs in front of $\sqrt{1-4w/s}$ in the expressions of $t$ and $u$ could also be chosen oppositely.
In our convention, for $s \to -\infty$, we have $u \to \infty$ while $t \to 0$, indicating nonzero $\Im \mathcal{M}$ in the $u$-channel at $s < 0$. On the other hand, the opposite case leads to  nonzero $\Im \mathcal{M}$ in the $t$-channel at $s < 0$.}
For amplitudes symmetric under $t \leftrightarrow u$, the presence of the square roots in \eq{eq:t_and_u} does not introduce any non-analyticity, since $t$ and $u$ always appear together in combinations such as $t + u$, $tu$, etc.\footnote{It indicates that any $t\leftrightarrow u$ symmetric amplitude is independent of the signs in front of $\sqrt{1-4w/s}$.} For this reason, the non-analytic parts continue to arise only from the real $s$ axis due to $s$- and $u$-channel particle exchanges.
They give rise to Im${\cal M}$ which, as in Sec.~\ref{sec:dispersion}, 
we expand in partial-waves:  
\begin{align}
\qquad
\mbox{Im}\mathcal{M}(s,w)&=
\sum_{J=\rm even}(2J+1)\rho^J(s)\,d^J_{\lam_{12},\lam_{43}}
\bigg(\sqrt{1-\frac{4w}{s}}\bigg)&&\mbox{for $s\geq 0$\,,\qquad}
\label{eq:pw_w_s}
\\
\qquad
\mbox{Im}\mathcal{M}(s,w)&=
\sum_{J}(2J+1)\rho^J(u)\,d^J_{\lam_{14},\lam_{23}}
\bigg(\frac{u-w}{u+w}\bigg)&&\mbox{for $u\geq 0$\,.\qquad}
\label{eq:pw_w_u}
\end{align}
The presence of the squared root in \eq{eq:pw_w_s} does not pose any problem  for $t\leftrightarrow u$ symmetric amplitudes, as it can be understood in the following way.
 The $t\leftrightarrow u$ symmetry requires that either the incoming or the outgoing particles are the same. Considering the first case, we have  $\lambda_{12}=0$, and consequently  that $J$ must be even in \eq{eq:pw_w_s}. 
In addition, $\lambda_{43}$ must also be even so, $d^J_{0,\lambda_{43}}$ is an even polynomial in its argument, which guarantees the absence of the square root in Eq.~\eqref{eq:pw_w_s}.

Taking $w$ fixed, we can write the following  dispersion relation for $\mathcal{M}(s,w)/{s^{1+k}}$
using \eq{eq:pw_w_s} and \eq{eq:pw_w_u}:
\begin{align}
\frac{1}{2\pi i}\oint_{C_0}ds' 
\frac{\mathcal{M}(s',w)}{s^{\prime 1+k}}
=&\Bigg\langle 
\frac{d^J_{\lambda_{12},\lambda_{43}}\Big(\sqrt{1-\frac{4w}{m^2}}\Big)}{m^{2 k}}
\Bigg \rangle
+(-1)^{k} \Bigg\langle 
\frac{ d^J_{\lambda_{14},\lambda_{23}}\Big(\frac{m^2-w}{m^2+w}\Big)}{m^{4k}}
(m^2+2w)(m^2+w)^{k-1}
\Bigg \rangle\,.
\label{eq:dispersion_rel_at_fixed_w}
\end{align}

\section{Explicit expressions of the smeared Wigner $d$-functions}
\label{appendix:explicit_smeared}

In this section, we present the explicit formulas for the smeared Wigner $d$-functions used in the main text. 

Those in Eq.~\eqref{eq:smeared_rrhh} for $\gamma^+\gamma^+\rightarrow h^+h^+$  are given by
\begingroup
\small
\begin{align}
A_s(m,J)
=
\int^0_{-|t|_{\rm max}} dt \, A(t)
\frac{d^J_{0,0}(1+2t/m^2)}{m^4}
,\quad 
A_{u}(m,J)
=
\int^0_{-|t|_{\rm max}} dt \, A(t)
\frac{m^2(-1)^J d^J_{3,3}(1+2t/m^2)}{
(m^2+t)^3},
\end{align}
\endgroup
corresponding to  the $s$- and $u$-channel respectively. 

The ones in Eqs.~\eqref{eq:smeared_rrrr},~\eqref{eq:smeared_hhhh}, and~\eqref{eq:smeared_rhrh} for $\gamma^+\gamma^+\rightarrow \gamma^+\gamma^+$, $h^+h^+\rightarrow h^+h^+$, and $\gamma^+ h^-\rightarrow \gamma^+ h^-$ are given respectively by
\begingroup
\small
\begin{gather}
B_{s}(m,J)=\int^{w_{\rm max}}_{0} dw \,
B(w)
\frac{d^J_{0,0}\Big(\sqrt{1-\frac{4w}{m^2}}\Big)}{m^4},
\\
B_{u}(m,J)=\int^{w_{\rm max}}_{0} dw \,
B(w)
\frac{d^J_{2,2}\Big(\frac{m^2-w}{m^2+w}\Big)}{m^8}(m^2+2w)(m^2+w),
\\
C_{s}(m,J)=\int^{0}_{-|t|_{\max}} dt \,
C(t)
\frac{d^J_{0,0}\big(1+\frac{2t}{m^2}\big)}{m^4}
,\quad
C_{u}(m,J)=\int^{0}_{-|t|_{\max}} dt \,
C(t)
\frac{m^2 d^J_{4,4}\big(1+\frac{2t}{m^2}\big)}{(m^2+t)^3},
\\
D_{s}(m,J)=\int^{0}_{-|t|_{\max}} dt \,
D(t)
\frac{d^J_{3,3}\big(1+\frac{2t}{m^2}\big)}{(m^2+t)^2}
,\quad
D_{u}(m,J)=\int^{0}_{-|t|_{\max}} dt \,
D(t)
\frac{m^2 d^J_{1,1}\big(1+\frac{2t}{m^2}\big)}{m^2(m^2+t)}\,.
\label{eq:D_smeared}
\end{gather}
\endgroup

In Eq.~\eqref{eq:smeared_rhrh2} for $\gamma^+h^+\rightarrow \gamma^- h^+$, we have defined
\begingroup
\small
\begin{gather}
E_s(m,J)
=
\int^0_{-|t|_{\rm max}} dt \, E(t)
\frac{(2m^2+t)d^J_{3,1}\big(1+\frac{2t}{m^2}\big)}{m^2
(m^2+t)^2}.
\end{gather}
\endgroup
Finally, the functions in Eq.~\eqref{eq:smeared_rhrh3} for $\gamma^+h^-\rightarrow \gamma^+ h^-$ are
\begingroup
\small
\begin{gather}
F_{s}(m,J)=\int^{0}_{-|t|_{\max}} dt \,
F(t)
\frac{d^J_{3,3}\big(1+\frac{2t}{m^2}\big)}{(m^2+t)^2}
,\quad
F_{u}(m,J)=\int^{0}_{-|t|_{\max}} dt \,
F(t)
\frac{m^2 d^J_{1,1}\big(1+\frac{2t}{m^2}\big)}{m^2(m^2+t)},
\end{gather}
\endgroup
which are the same as Eq.~\eqref{eq:D_smeared}, except for the smearing function.

\section{Positivity of smeared Wigner $d$-functions}
\label{appendix:positivity_of_smeared_d}
\numberwithin{equation}{section}
\setcounter{equation}{0}

\begin{figure}[t!]
\centering
\includegraphics[scale=0.65]{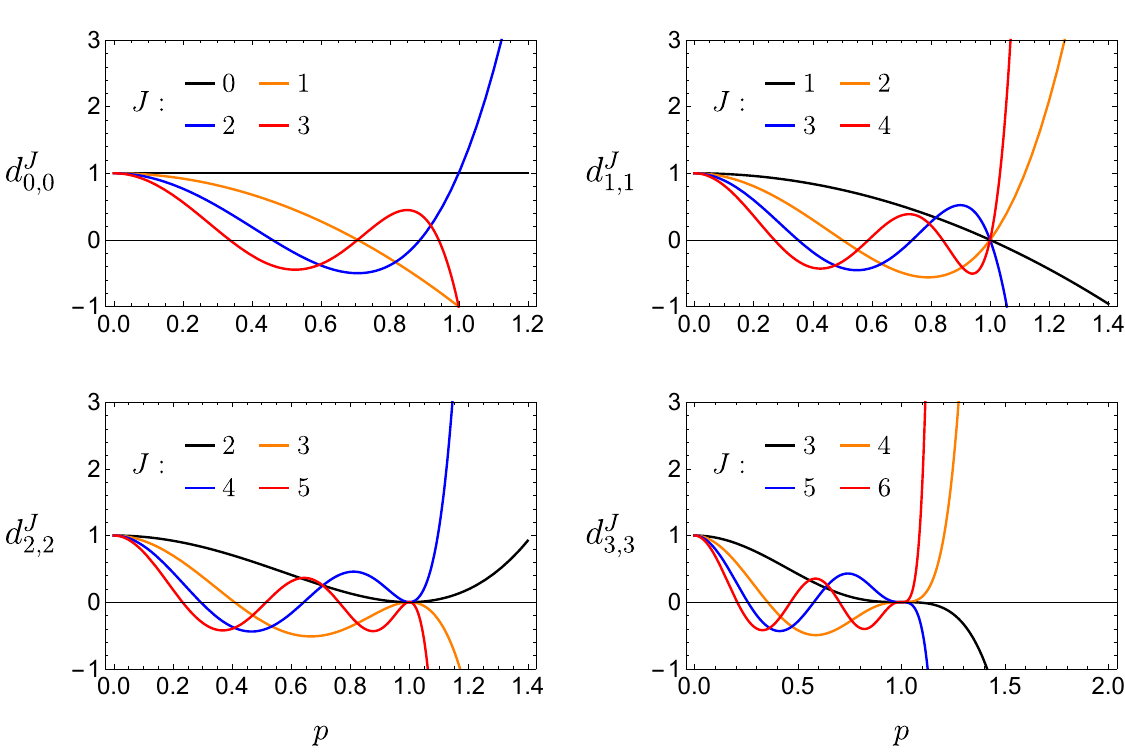}\qquad\;\;
\caption{Winger $d$-functions $d^J_{\lambda,\lambda}(1-2p^2)$.
}
\label{fig:d_0123}
\end{figure}
In this section we discuss in detail how to choose the smearing functions and determine $|t|_{\rm max}$
such that  $B_{s,u}(m,J), C_{s,u}(m,J), D_{s,u}(m,J)$ and $F_{s,u}(m,J)$ are positive.
All these functions appear in elastic processes so their corresponding  Winger $d$-functions 
have $\lambda_{12}=\lambda_{34}\equiv \lambda$. 

We first consider $C_{s,u}(m,J), D_{s,u}(m,J)$ and $F_{s,u}(m,J)$, that can be written  in a generic way as
\begin{align}
\mathcal{S}^J_{\lam,\lam}=
\int^{p_{\rm max}}_0 dp\, 2p\, S(p) \,d^J_{\lam,\lam}(1-2p^2) \quad \mbox{with} 
\quad p=\sqrt{-t/m^2}\,.
\label{eq:smear0}
\end{align}
Notice that  in $S(p)$ we incorporate not only the  smearing function  but also  possible powers of  $m^2/(m^2+t)=1/(1-p^2)$.

We begin by deriving a necessary condition for the positivity of  Eq.~\eqref{eq:smear0} in the limit $J\to\infty$ at fixed $m^2$. 
As shown in  Fig.~\ref{fig:d_0123},   the Winger $d$-functions   monotonically increases or decreases for $p > 1$, leading to $d^J_{\lam,\lam} \rightarrow (-1)^{J} \infty$ as $p \rightarrow \infty$, except for $d^0_{0,0} = 1$.\footnote{We will deduce this limit  in \eq{eq:d_asym}.} 
 Furthermore,  the divergence toward $\pm\infty$ becomes stronger  as $J$ increases; therefore, it cannot be tamed by any smearing function in the $J\to\infty$ limit.
This forces us to impose $p_{\text{max}} \leq 1$ in order to ensure that \eq{eq:smear0} remains positive as $J\to\infty$.\footnote{$C_s(m,J)$ escapes this logic since only even $J$ is allowed, and therefore its positive sign for $p\gg 1$ is fixed; with a proper smearing function, it can be made always positive.}

Fig.~\ref{fig:d_0123} also shows that the Wigner $d$-functions with identical sub-indices are positive 
around $p \sim 0$ for all $J$, while oscillates as  $p$ approaches $1$.
Thus, the smearing function must be chosen to enhance the region $p \sim 0$ and  suppress  the oscillatory behavior. 
This can be achieved by setting the smearing function  proportional to
\begin{align}
S(p)\propto (1-p/p_{\rm max})^r = (1-\sqrt{-t/|t|_{\rm max}})^r 
\quad \mbox{with}\quad r\geq 0,
\label{eq:smearing_propto}
\end{align}
where $p/p_{\rm max}$ is introduced to prevent the smearing function 
from depending on $m$. In the main text, we set $r=5$, which is sufficient to suppress oscillatory contributions. We note that although many alternatives of the smearing function could satisfy the positivity conditions discussed so far, they do not result in a significant difference in the final bound~\cite{Caron-Huot:2022jli}.

We can proceed in the same way for 
$B_{s,u}(m,J)$ in which the  smearing is performed over $w$ instead of $t$.  In this case,  a few remarks are in order.
For $B_{s}(m,J)$ we observe that the Wigner $d$-function involved monotonically increases (decreases) for even (odd) $J/2>0$
in the region $w> m^2/4$.\footnote{For $p=\sqrt{4w/m^2}$, we have
$d^J_{0,0}(\sqrt{1-p^2})\propto (-1)^{J/2}(p-1)$ near $p=1$ for even $J>0$.}
Thus, we must require
\begin{align}
w_{\rm max}\leq m^2/4 \quad \mbox{for $B_s(m,J)$},
\label{eq:wmax_cond}
\end{align}
to ensure positivity for all  $J$. 
On the other hand, we find that  for $B_u(m,J)$ one can always find a smearing function that guarantees positivity. 
This is because the argument of the Wigner $d$-function is $(m^2-w)/(m^2+w)$ that is always between $1$ and $-1$
where  the Wigner $d$-function does not diverge for any $J$.

\subsection{Maximal  $|t|_{\rm max}$  as a function of $J$}
\label{appendix:t_max}

In the previous section, we have seen that   a generic bound to have Eq.~\eqref{eq:smear0} positive for all $J$ 
corresponds to  $p_{\rm max}=\sqrt{|t|_{\rm max}/m^2} \leq 1$. This fixes  $|t|_{\rm max} = {\rm Min}[M_s, M_u]$, where we recall that $M_{s,u}$ are the masses of the  $s$- and $u$-channel lightest UV state.

Nevertheless, one can consider the possibility of taking larger values of $|t|_{\rm max}$, as this would lead to stronger bounds. We aim here to identify the largest $J$ for which we can take $|t|_{\mathrm{max}} = M_J^2$ without losing positivity,
where  $M_J$ denotes the mass of the lightest UV state with spin $J$. 
As shown in the main text, this will determine the spin of the  states that must appear at the causal cutoff
$\Lambda_{\mathrm{caus}}$, providing crucial information about the UV completion.

The above goal requires a detail inspection of smearing each Wigner $d$-function as a function of $J$, starting from its lowest possible value. We obtain the following result. Taking $|t|_{\mathrm{max}}=M_J^2$, the  values of $J$ are
\begin{itemize}
\item $J\geq 2$ for $B_s\geq0$ \,,
\item $J\geq 3$ for $B_u\geq0,D_u\geq0,F_u\geq0$  \,,
\item $J\geq 4$ for $C_s\geq0,D_s\geq0,F_s\geq0$  \,,
\end{itemize}
with no restriction on $J$ for satisfying $C_s \ge 0$. In the rest of the section, we provide a detailed explanation of the cases of $B_s(m,J)$ and $F_u(m,J)$, which determine the  $J$ values in \eq{maxt} and \eq{eq:t_M3}, respectively. These are essential for interpreting the bounds~\eq{eq:bound_kkg} and~\eq{eq:bound_k1}. The remaining cases  can be straightforwardly obtained by following the argument presented here.

$B_s(m,J)$ includes $d^J_{0,0}$ allowing, in principle, $J=0,2,4,...$.\footnote{Recall that $B_s(m,J)$ arise from the $s$-channel of $\gamma^+\gamma^+\rightarrow \gamma^+\gamma^+$, restricting $J$ to even values.} For $J=0$,
$d^0_{0,0}$ is constant, so it is easy to guarantee the positivity of $B_s(m,0)$ for any $w_{\rm max}$. Next, we consider  $d^2_{0,0}$. It  decreases for $\omega_{\rm max}>m^2/4$, requiring then to take
$w_{\rm max}=M^2_{J=2}/4$, or equivalently, $|t|_{\rm max}=M^2_{J=2}$
to have $B_s(m,2)\geq 0$. 
The UV state of mass  $M_{J=2}$ couples to $\gamma^+\gamma^+$ and $h^-h^-$.

For  $F_{u}(m,J)$ (with the same structure as $D_{u}(m,J)$), involving $d^J_{1,1}/(m^2+t)$, we can have $J=1,2,3,...\,$. The proportionality,
\begin{align}
d^J_{\lam,\lam}(1-2p^2)\propto (1-p^2)^{|\lam|} P^{(0,|\lam|)}_{J-|\lam|}(1-2p^2),
\end{align}
with $P^{(\alpha,\beta)}_{n}$ the Jacobi polynomial,  guarantees that $d^1_{1,1}/(m^2+t)$ is  constant, and $F_u(m,1)\geq 0$  for any $|t|_{\rm max}$. On the other hand, $d^2_{1,1}/(m^2+t)$ decreases monotonically to  negative values for $|t|>m^2$. In spite of this, by taking the smearing function to involve a Bessel function, we can cancel the negative contribution and obtain $F_u(m,2)\geq 0$. 
In contrast, the Bessel function cancels the positive contribution of $d^3_{1,1}/(m^2+t)$ for $|t|>m^2$, determining that   $|t|_{\rm max}$ cannot be larger than  $M^2_{J=3}$
in order to have $F_u(m^2,3)\geq 0$. The UV state of mass $M_{J=3}$ couples to $\gamma^+ h^+$ and $\gamma^-h^+$.

\subsection{High-$J$ asymptotics of the smeared Wigner $d$-functions}
\label{appendix:high_J_asym}
\numberwithin{equation}{section}

To confirm the positivity of $B_{s,u}(m,J)$, $C_{s,u}(m,J)$, $D_{s,u}(m,J)$ and $F_{s,u}(m,J)$, and also the validity of Eqs.~\eqref{eq:conditions} and~\eqref{eq:E_F_condition} for all $J$, we must understand the high-$J$ asymptotics of the smeared Wigner $d$-functions as numerical evaluation cannot access the $J \to \infty$ limit.

We first consider the aymptotics of $S^J_{\lam,\lam'}$ with different sub-indices from \eq{eq:smear0} for any $p_{\rm max}\leq 1$, applicable to the smeared Wigner $d$-functions with the variable $t$. For the lower $p_{\rm max}< 1$, the smearing function with $(1-p/p_{\rm max})^r$ suppresses the oscillatory contributions more strongly.

Therefore, it suffices to evaluate the asymptotics of $S^J_{\lam,\lam'}$ only at $p_{\rm max}=1$, leading to our target: 
\begin{align}
\mathcal{A}_{\lam,\lam'}(J)\equiv
\lim_{J\to \infty}
\mathcal{S}^J_{\lam,\lam'}
=
\lim_{J\to \infty}
\int^\pi_0 \frac{\sin\theta d\theta}{2} S(\sin\sfrac{\theta}{2}) d^J_{\lam,\lam'}(\cos\theta),
\label{eq:smear1}
\end{align}
where we have changed  variables,  $\cos\theta\equiv 1-2p^2$. In the high-$J$ limit, the Wigner $d$-function becomes rapidly oscillatory, but the stationary phase method~\cite{BleisteinHandelsman1986,SteinShakarchi2005} shows that bulk contributions cancel while the endpoint contributions remain. This allows us to use the asymptotic expressions of the Wigner $d$-function instead of the exact form:\footnote{In this work, we adopt the convention of the $d$-function in Ref.~\cite{Rose:1995}.}
\begin{align}
d^J_{\lam,\lam'}(\theta)
\sim 
\left\{
\begin{array}{ll}
(-1)^{\frac{\lam_-+|\lam_-|}{2}} J_{|\lam_-|}(J \theta)
&\mbox{for $\theta \ll 1$},
\\[8pt]
(-1)^{\frac{-|\lam_-|+|\lam_+|-2J }{2}} J_{|\lam_+|}
\big[J (\pi-\theta)\big]
& \mbox{for $\pi-\theta \ll 1$}
\end{array}
\right.,
\label{eq:d_asym}
\end{align}
at $J \gg 1$ where $\lam_\pm=\lam\pm \lam'$ and $J_{\lam}$ is the Bessel function of the first kind. This is because the two asymptotics match the Wigner $d$-function at the endpoints, $\theta=0$ and $\pi$, respectively, while their magnitude is suppressed as $1/\sqrt{J}$ in the bulk, enabling us to rewrite Eq.~\eqref{eq:smear1} as
\begin{align}
\mathcal{A}_{\lam,\lam'}(J)
=&
\lim_{J\to \infty}
\bigg\{
(-1)^{\frac{\lam_-+|\lam_-|}{2}}
\int^a_0 \frac{\sin\theta d\theta}{2} S(\sin\sfrac{\theta}{2})
J_{|\lam_-|}(J \theta)
\nonumber
\\
&+(-1)^{\frac{-|\lam_-|+|\lam_+|-2J }{2}}
\int^\pi_b \frac{\sin\theta d\theta}{2} S(\sin\sfrac{\theta}{2})
J_{|\lam_+|}\big[J (\pi-\theta)\big]
\bigg\},
\label{eq:smear2}
\end{align}
in terms of arbitrarily finite cutoffs, $a>0$ and $b<\pi$.

Due to the endpoint dominance, we can use the Taylor expansion of the smearing function with the sine factor near the endpoints. The smearing function that we consider is
\begin{align}
S(\sin\sfrac{\theta}{2})\equiv (\sin\sfrac{\theta}{2})^{\alpha}\widetilde{S}(\sin\sfrac{\theta}{2}) \quad \mbox{with}\quad 0\leq \alpha <1,
\end{align}
which is non-analytic at $\theta=0$ for $\alpha>0$. Expanding this as 
\begin{align}
&\frac{\sin\theta}{2}S(\sin\sfrac{\theta}{2})
=
\frac{\theta^\alpha}{2^{1+\alpha}}
\bigg[\theta \widetilde{S}(0)+\frac{\theta^2}{2}\widetilde{S}'(0)+\cdots\bigg]
&&\mbox{for $\theta \ll 1$},
\label{eq:taylor_exp1}
\\
&\frac{\sin\theta}{2}S(\sin\sfrac{\theta}{2})
=
\frac12\bigg[\theta \widetilde{S}(1)-\frac{\theta^2}{24}\big[(4+3\alpha)\widetilde{S}(1)+\widetilde{S}'(1)\big]+\cdots\bigg]
&&\mbox{for $\pi-\theta \ll 1$}, 
\label{eq:taylor_exp2}
\end{align}
at each endpoint, we can get the asymptotics by putting the expansions above into \eq{eq:smear2}:
\begin{align}
\mathcal{A}_{\lam,\lam'}(J)
&= (-1)^{\frac{\lam_-+|\lam_-|}{2}}
\Bigg\{
\frac{\Gamma[(2+|\lam_-|+\alpha)/2]}{J^{2+\alpha}\Gamma[(|\lam_-|-\alpha)/2]}\widetilde{S}^{(0)}(0)
+
\frac{\Gamma[(3+|\lam_-|+\alpha)/2]}{J^{3+\alpha}\Gamma[(|\lam_-|-1-\alpha)/2]}\widetilde{S}^{(1)}(0)
+\cdots
\Bigg\}
\nonumber
\\
&\quad +(-1)^{\frac{-\lam_-+|\lam_+|-2J}{2}}
\Bigg\{
\frac{|\lam_+|}{2J^2}\widetilde{S}^{(0)}(1)
-\frac{|\lam_+|(|\lam_+|^2-4)}{48 J^4}
\big[(4+3\alpha)\widetilde{S}^{(0)}(1)+3\widetilde{S}^{(1)}(1)\big]
+\cdots
\Bigg\},
\end{align}
Here, we used the analytic integral,
\begin{equation}
\lim_{J\to \infty}\int_{0}^{a} \theta^{1+n} J_{\lam}(J\theta)\,d\theta
= \frac{2^{1+n}\Gamma\big(1+\frac{\lam+n}{2}\big)}{J^{2+n}\Gamma\big(\frac{\lam-n}{2}\big)}+\mbox{$a$-dependent oscillatory terms},
\label{eq:bessel-infty}
\end{equation}
where we neglected the oscillatory terms, ensured by the stationary phase method.

Similarly to the case of smeared Wigner $d$-functions with the variable $t$, we can derive the asymptotics of those with the variable $w$. Taking the critical case $w_{\rm max}=m^2/4$ in \eq{eq:wmax_cond}, the asymptotics involving $d^J_{\lam,\lam}(\sqrt{1-4w/m^2})$ can be given by
\begin{align}
\int^{\pi/2}_0 \sin2\theta \,d\theta\, S(\sin\theta) \,
d^J_{\lam,\lam}(\cos\theta),
\label{eq:smeared_d_fixed_w1}
\end{align}
where we use $\sqrt{1-4w/m^2}=\cos\theta$. Note that the integration region is restricted to $\pi/2$, allowing only the endpoint $\theta=0$ contribution to be dominant due to the $1/\sqrt{J}$ supression of the Winger $d$-function at bulk.

As mentioned before, the argument $(m^2-w)/(m^2+w)$ of the Wigner $d$-function is put only between $-1$ and $1$, allowing to make always the relevant smeared ones positive via a suitable choice of the smearing function. However, as shown in Eq.~\eqref{eq:wmax_cond}, our choice making $B_s(m,J)$ positive for $w_{\rm max}\leq m^2/4$ leads $B_u(m,J)$ to be negative for a specific $J$ value. It is easy to see where the negativity arise from by writing the smeared one as
\begin{align}
\int^{\Theta_{\rm max}}_0 \frac{2\sin\theta\, d\theta}{(1+\cos\theta)^2}  
S\bigg(\frac{1-\cos\theta}{\sin\theta}\bigg) \,
d^J_{\lam,\lam}(\cos\theta),
\label{eq:smeared_d_fixed_w2}
\end{align}
where we use $(m^2-w)/(m^2+w)=\cos\theta$ with $\Theta_{\rm max}=
\cos^{-1}[(m^2-w_{\rm max})/(m^2+w_{\rm max})]$. For the function $S$ adopting the smearing function in Eq.~\eqref{eq:smearing_B} is proportional to
\begin{align}
S\bigg(\frac{1-\cos\theta}{\sin\theta}\bigg)\propto 4\bigg(1-\frac{1-\cos\theta}{\Delta_{\rm max} \sin\theta}\bigg)^5 \quad \mbox{with} \quad 
\Delta_{\rm max}=\frac{1-\cos\Theta_{\rm max}}{\sin\Theta_{\rm max}},
\end{align}
which vanishes at $\theta=\Theta_{\rm max}$. This function cannot cancels the monotonic growth of the factor $\sin\theta/(1+\cos\theta)^2$ for $\theta \rightarrow \pi$ and moreover, the extra factor $(m^2+2w)(m^2+w)^{k-1}$ is replaced to involve an additional power of $1/(1+\cos\theta)$ in the integral in Eq.~\eqref{eq:smeared_d_fixed_w2}. Since $d^J_{\lam,\lam}$ is oscillatory near $\theta=\pi$, it leads us to lose the positivity for all $J$.

\bibliographystyle{utphys}
\bibliography{ref}

\providecommand{\noopsort}[1]{}\providecommand{\singleletter}[1]{#1}%
\providecommand{\href}[2]{#2}\begingroup\raggedright\begin{thebibliography}{10}

\bibitem{Kruczenski:2022lot}
M.~Kruczenski, J.~Penedones, and B.~C. van Rees, ``{Snowmass White Paper: S-matrix Bootstrap},'' \href{https://arxiv.org/abs/2203.02421}{{\ttfamily arXiv:2203.02421 [hep-th]}}.

\bibitem{Paulos:2016but}
M.~F. Paulos, J.~Penedones, J.~Toledo, B.~C. van Rees, and P.~Vieira, ``{The S-matrix bootstrap II: two dimensional amplitudes},'' \href{https://dx.doi.org/10.1007/JHEP11(2017)143}{{\em JHEP} {\bfseries 11} (2017) 143}, \href{https://arxiv.org/abs/1607.06110}{{\ttfamily arXiv:1607.06110 [hep-th]}}.

\bibitem{Paulos:2017fhb}
M.~F. Paulos, J.~Penedones, J.~Toledo, B.~C. van Rees, and P.~Vieira, ``{The S-matrix bootstrap. Part III: higher dimensional amplitudes},'' \href{https://dx.doi.org/10.1007/JHEP12(2019)040}{{\em JHEP} {\bfseries 12} (2019) 040}, \href{https://arxiv.org/abs/1708.06765}{{\ttfamily arXiv:1708.06765 [hep-th]}}.

\bibitem{Guerrieri:2018uew}
A.~L. Guerrieri, J.~Penedones, and P.~Vieira, ``{Bootstrapping QCD Using Pion Scattering Amplitudes},'' \href{https://dx.doi.org/10.1103/PhysRevLett.122.241604}{{\em Phys. Rev. Lett.} {\bfseries 122} no.~24, (2019) 241604}, \href{https://arxiv.org/abs/1810.12849}{{\ttfamily arXiv:1810.12849 [hep-th]}}.

\bibitem{Doroud:2018szp}
N.~Doroud and J.~Elias~Mir{\'o}, ``{S-matrix bootstrap for resonances},'' \href{https://dx.doi.org/10.1007/JHEP09(2018)052}{{\em JHEP} {\bfseries 09} (2018) 052}, \href{https://arxiv.org/abs/1804.04376}{{\ttfamily arXiv:1804.04376 [hep-th]}}.

\bibitem{deRham:2018qqo}
C.~de~Rham, S.~Melville, A.~J. Tolley, and S.-Y. Zhou, ``{Positivity Bounds for Massive Spin-1 and Spin-2 Fields},'' \href{https://dx.doi.org/10.1007/JHEP03(2019)182}{{\em JHEP} {\bfseries 03} (2019) 182}, \href{https://arxiv.org/abs/1804.10624}{{\ttfamily arXiv:1804.10624 [hep-th]}}.

\bibitem{Zhang:2018shp}
C.~Zhang and S.-Y. Zhou, ``{Positivity bounds on vector boson scattering at the LHC},'' \href{https://dx.doi.org/10.1103/PhysRevD.100.095003}{{\em Phys. Rev. D} {\bfseries 100} no.~9, (2019) 095003}, \href{https://arxiv.org/abs/1808.00010}{{\ttfamily arXiv:1808.00010 [hep-ph]}}.

\bibitem{Bellazzini:2019bzh}
B.~Bellazzini, F.~Riva, J.~Serra, and F.~Sgarlata, ``{Massive Higher Spins: Effective Theory and Consistency},'' \href{https://dx.doi.org/10.1007/JHEP10(2019)189}{{\em JHEP} {\bfseries 10} (2019) 189}, \href{https://arxiv.org/abs/1903.08664}{{\ttfamily arXiv:1903.08664 [hep-th]}}.

\bibitem{Remmen:2019cyz}
G.~N. Remmen and N.~L. Rodd, ``{Consistency of the Standard Model Effective Field Theory},'' \href{https://dx.doi.org/10.1007/JHEP12(2019)032}{{\em JHEP} {\bfseries 12} (2019) 032}, \href{https://arxiv.org/abs/1908.09845}{{\ttfamily arXiv:1908.09845 [hep-ph]}}.

\bibitem{Bellazzini:2019xts}
B.~Bellazzini, M.~Lewandowski, and J.~Serra, ``{Positivity of Amplitudes, Weak Gravity Conjecture, and Modified Gravity},'' \href{https://dx.doi.org/10.1103/PhysRevLett.123.251103}{{\em Phys. Rev. Lett.} {\bfseries 123} no.~25, (2019) 251103}, \href{https://arxiv.org/abs/1902.03250}{{\ttfamily arXiv:1902.03250 [hep-th]}}.

\bibitem{Huang:2020nqy}
Y.-t. Huang, J.-Y. Liu, L.~Rodina, and Y.~Wang, ``{Carving out the Space of Open-String S-matrix},'' \href{https://dx.doi.org/10.1007/JHEP04(2021)195}{{\em JHEP} {\bfseries 04} (2021) 195}, \href{https://arxiv.org/abs/2008.02293}{{\ttfamily arXiv:2008.02293 [hep-th]}}.

\bibitem{Sinha:2020win}
A.~Sinha and A.~Zahed, ``{Crossing Symmetric Dispersion Relations in Quantum Field Theories},'' \href{https://dx.doi.org/10.1103/PhysRevLett.126.181601}{{\em Phys. Rev. Lett.} {\bfseries 126} no.~18, (2021) 181601}, \href{https://arxiv.org/abs/2012.04877}{{\ttfamily arXiv:2012.04877 [hep-th]}}.

\bibitem{Bellazzini:2020cot}
B.~Bellazzini, J.~Elias~Mir{\'o}, R.~Rattazzi, M.~Riembau, and F.~Riva, ``{Positive moments for scattering amplitudes},'' \href{https://dx.doi.org/10.1103/PhysRevD.104.036006}{{\em Phys. Rev. D} {\bfseries 104} no.~3, (2021) 036006}, \href{https://arxiv.org/abs/2011.00037}{{\ttfamily arXiv:2011.00037 [hep-th]}}.

\bibitem{Tolley:2020gtv}
A.~J. Tolley, Z.-Y. Wang, and S.-Y. Zhou, ``{New positivity bounds from full crossing symmetry},'' \href{https://dx.doi.org/10.1007/JHEP05(2021)255}{{\em JHEP} {\bfseries 05} (2021) 255}, \href{https://arxiv.org/abs/2011.02400}{{\ttfamily arXiv:2011.02400 [hep-th]}}.

\bibitem{Hebbar:2020ukp}
A.~Hebbar, D.~Karateev, and J.~Penedones, ``{Spinning S-matrix bootstrap in 4d},'' \href{https://dx.doi.org/10.1007/JHEP01(2022)060}{{\em JHEP} {\bfseries 01} (2022) 060}, \href{https://arxiv.org/abs/2011.11708}{{\ttfamily arXiv:2011.11708 [hep-th]}}.

\bibitem{Arkani-Hamed:2020blm}
N.~Arkani-Hamed, T.-C. Huang, and Y.-t. Huang, ``{The EFT-Hedron},'' \href{https://dx.doi.org/10.1007/JHEP05(2021)259}{{\em JHEP} {\bfseries 05} (2021) 259}, \href{https://arxiv.org/abs/2012.15849}{{\ttfamily arXiv:2012.15849 [hep-th]}}.

\bibitem{Caron-Huot:2020cmc}
S.~Caron-Huot and V.~Van~Duong, ``{Extremal Effective Field Theories},'' \href{https://dx.doi.org/10.1007/JHEP05(2021)280}{{\em JHEP} {\bfseries 05} (2021) 280}, \href{https://arxiv.org/abs/2011.02957}{{\ttfamily arXiv:2011.02957 [hep-th]}}.

\bibitem{Caron-Huot:2021rmr}
S.~Caron-Huot, D.~Mazac, L.~Rastelli, and D.~Simmons-Duffin, ``{Sharp boundaries for the swampland},'' \href{https://dx.doi.org/10.1007/JHEP07(2021)110}{{\em JHEP} {\bfseries 07} (2021) 110}, \href{https://arxiv.org/abs/2102.08951}{{\ttfamily arXiv:2102.08951 [hep-th]}}.

\bibitem{Bern:2021ppb}
Z.~Bern, D.~Kosmopoulos, and A.~Zhiboedov, ``{Gravitational effective field theory islands, low-spin dominance, and the four-graviton amplitude},'' \href{https://dx.doi.org/10.1088/1751-8121/ac0e51}{{\em J. Phys. A} {\bfseries 54} no.~34, (2021) 344002}, \href{https://arxiv.org/abs/2103.12728}{{\ttfamily arXiv:2103.12728 [hep-th]}}.

\bibitem{Chiang:2021ziz}
L.-Y. Chiang, Y.-t. Huang, W.~Li, L.~Rodina, and H.-C. Weng, ``{Into the EFThedron and UV constraints from IR consistency},'' \href{https://dx.doi.org/10.1007/JHEP03(2022)063}{{\em JHEP} {\bfseries 03} (2022) 063}, \href{https://arxiv.org/abs/2105.02862}{{\ttfamily arXiv:2105.02862 [hep-th]}}.

\bibitem{Henriksson:2021ymi}
J.~Henriksson, B.~McPeak, F.~Russo, and A.~Vichi, ``{Rigorous bounds on light-by-light scattering},'' \href{https://dx.doi.org/10.1007/JHEP06(2022)158}{{\em JHEP} {\bfseries 06} (2022) 158}, \href{https://arxiv.org/abs/2107.13009}{{\ttfamily arXiv:2107.13009 [hep-th]}}.

\bibitem{Zahed:2021fkp}
A.~Zahed, ``{Positivity and geometric function theory constraints on pion scattering},'' \href{https://dx.doi.org/10.1007/JHEP12(2021)036}{{\em JHEP} {\bfseries 12} (2021) 036}, \href{https://arxiv.org/abs/2108.10355}{{\ttfamily arXiv:2108.10355 [hep-th]}}.

\bibitem{Alvarez:2021kpq}
B.~Alvarez, J.~Bijnens, and M.~Sj{\"o}, ``{NNLO positivity bounds on chiral perturbation theory for a general number of flavours},'' \href{https://dx.doi.org/10.1007/JHEP03(2022)159}{{\em JHEP} {\bfseries 2022} (2022) 159}, \href{https://arxiv.org/abs/2112.04253}{{\ttfamily arXiv:2112.04253 [hep-ph]}}.

\bibitem{EliasMiro:2022xaa}
J.~Elias~Miro, A.~Guerrieri, and M.~A. Gumus, ``{Bridging positivity and S-matrix bootstrap bounds},'' \href{https://dx.doi.org/10.1007/JHEP05(2023)001}{{\em JHEP} {\bfseries 05} (2023) 001}, \href{https://arxiv.org/abs/2210.01502}{{\ttfamily arXiv:2210.01502 [hep-th]}}.

\bibitem{Serra:2022pzl}
F.~Serra, J.~Serra, E.~Trincherini, and L.~G. Trombetta, ``{Causality constraints on black holes beyond GR},'' \href{https://dx.doi.org/10.1007/JHEP08(2022)157}{{\em JHEP} {\bfseries 08} (2022) 157}, \href{https://arxiv.org/abs/2205.08551}{{\ttfamily arXiv:2205.08551 [hep-th]}}.

\bibitem{Caron-Huot:2022ugt}
S.~Caron-Huot, Y.-Z. Li, J.~Parra-Martinez, and D.~Simmons-Duffin, ``{Causality constraints on corrections to Einstein gravity},'' \href{https://dx.doi.org/10.1007/JHEP05(2023)122}{{\em JHEP} {\bfseries 05} (2023) 122}, \href{https://arxiv.org/abs/2201.06602}{{\ttfamily arXiv:2201.06602 [hep-th]}}.

\bibitem{Caron-Huot:2022jli}
S.~Caron-Huot, Y.-Z. Li, J.~Parra-Martinez, and D.~Simmons-Duffin, ``{Graviton partial waves and causality in higher dimensions},'' \href{https://dx.doi.org/10.1103/PhysRevD.108.026007}{{\em Phys. Rev. D} {\bfseries 108} no.~2, (2023) 026007}, \href{https://arxiv.org/abs/2205.01495}{{\ttfamily arXiv:2205.01495 [hep-th]}}.

\bibitem{Henriksson:2022oeu}
J.~Henriksson, B.~McPeak, F.~Russo, and A.~Vichi, ``{Bounding violations of the weak gravity conjecture},'' \href{https://dx.doi.org/10.1007/JHEP08(2022)184}{{\em JHEP} {\bfseries 08} (2022) 184}, \href{https://arxiv.org/abs/2203.08164}{{\ttfamily arXiv:2203.08164 [hep-th]}}.

\bibitem{Albert:2022oes}
J.~Albert and L.~Rastelli, ``{Bootstrapping pions at large N},'' \href{https://dx.doi.org/10.1007/JHEP08(2022)151}{{\em JHEP} {\bfseries 08} (2022) 151}, \href{https://arxiv.org/abs/2203.11950}{{\ttfamily arXiv:2203.11950 [hep-th]}}.

\bibitem{Fernandez:2022kzi}
C.~Fernandez, A.~Pomarol, F.~Riva, and F.~Sciotti, ``{Cornering large-N$_{c}$ QCD with positivity bounds},'' \href{https://dx.doi.org/10.1007/JHEP06(2023)094}{{\em JHEP} {\bfseries 06} (2023) 094}, \href{https://arxiv.org/abs/2211.12488}{{\ttfamily arXiv:2211.12488 [hep-th]}}.

\bibitem{Acanfora:2023axz}
F.~Acanfora, A.~Guerrieri, K.~H{\"a}ring, and D.~Karateev, ``{Bounds on scattering of neutral Goldstones},'' \href{https://dx.doi.org/10.1007/JHEP03(2024)028}{{\em JHEP} {\bfseries 03} (2024) 028}, \href{https://arxiv.org/abs/2310.06027}{{\ttfamily arXiv:2310.06027 [hep-th]}}.

\bibitem{He:2023lyy}
Y.~He and M.~Kruczenski, ``{Bootstrapping gauge theories},'' \href{https://dx.doi.org/10.1103/PhysRevLett.133.191601}{{\em Phys. Rev. Lett.} {\bfseries 133} (2024) 191601}, \href{https://arxiv.org/abs/2309.12402}{{\ttfamily arXiv:2309.12402 [hep-th]}}.

\bibitem{Guerrieri:2023qbg}
A.~L. Guerrieri, A.~Hebbar, and B.~C. van Rees, ``{Constraining Glueball Couplings},'' \href{https://arxiv.org/abs/2312.00127}{{\ttfamily arXiv:2312.00127 [hep-th]}}.

\bibitem{Albert:2023jtd}
J.~Albert and L.~Rastelli, ``{Bootstrapping pions at large N. Part II. Background gauge fields and the chiral anomaly},'' \href{https://dx.doi.org/10.1007/JHEP09(2024)039}{{\em JHEP} {\bfseries 09} (2024) 039}, \href{https://arxiv.org/abs/2307.01246}{{\ttfamily arXiv:2307.01246 [hep-th]}}.

\bibitem{Ma:2023vgc}
T.~Ma, A.~Pomarol, and F.~Sciotti, ``{Bootstrapping the chiral anomaly at large N$_{c}$},'' \href{https://dx.doi.org/10.1007/JHEP11(2023)176}{{\em JHEP} {\bfseries 11} (2023) 176}, \href{https://arxiv.org/abs/2307.04729}{{\ttfamily arXiv:2307.04729 [hep-th]}}.

\bibitem{Bellazzini:2023nqj}
B.~Bellazzini, G.~Isabella, S.~Ricossa, and F.~Riva, ``{Massive gravity is not positive},'' \href{https://dx.doi.org/10.1103/PhysRevD.109.024051}{{\em Phys. Rev. D} {\bfseries 109} no.~2, (2024) 024051}, \href{https://arxiv.org/abs/2304.02550}{{\ttfamily arXiv:2304.02550 [hep-th]}}.

\bibitem{Li:2023qzs}
Y.-Z. Li, ``{Effective field theory bootstrap, large-N {\ensuremath{\chi}}PT and holographic QCD},'' \href{https://dx.doi.org/10.1007/JHEP01(2024)072}{{\em JHEP} {\bfseries 01} (2024) 072}, \href{https://arxiv.org/abs/2310.09698}{{\ttfamily arXiv:2310.09698 [hep-th]}}.

\bibitem{Albert:2023seb}
J.~Albert, J.~Henriksson, L.~Rastelli, and A.~Vichi, ``{Bootstrapping mesons at large N: Regge trajectory from spin-two maximization},'' \href{https://dx.doi.org/10.1007/JHEP09(2024)172}{{\em JHEP} {\bfseries 09} (2024) 172}, \href{https://arxiv.org/abs/2312.15013}{{\ttfamily arXiv:2312.15013 [hep-th]}}.

\bibitem{Hong:2023zgm}
D.-Y. Hong, Z.-H. Wang, and S.-Y. Zhou, ``{Causality bounds on scalar-tensor EFTs},'' \href{https://dx.doi.org/10.1007/JHEP10(2023)135}{{\em JHEP} {\bfseries 10} (2023) 135}, \href{https://arxiv.org/abs/2304.01259}{{\ttfamily arXiv:2304.01259 [hep-th]}}.

\bibitem{He:2024nwd}
Y.~He and M.~Kruczenski, ``{Gauge Theory Bootstrap: Pion amplitudes and low energy parameters},'' \href{https://arxiv.org/abs/2403.10772}{{\ttfamily arXiv:2403.10772 [hep-th]}}.

\bibitem{Guerrieri:2024jkn}
A.~Guerrieri, K.~H{\"a}ring, and N.~Su, ``{From data to the analytic S-matrix: A Bootstrap fit of the pion scattering amplitude},'' \href{https://arxiv.org/abs/2410.23333}{{\ttfamily arXiv:2410.23333 [hep-th]}}.

\bibitem{Beadle:2024hqg}
C.~Beadle, G.~Isabella, D.~Perrone, S.~Ricossa, F.~Riva, and F.~Serra, ``{Non-forward UV/IR relations},'' \href{https://dx.doi.org/10.1007/JHEP08(2025)188}{{\em JHEP} {\bfseries 08} (2025) 188}, \href{https://arxiv.org/abs/2407.02346}{{\ttfamily arXiv:2407.02346 [hep-th]}}.

\bibitem{Bertucci:2024qzt}
F.~Bertucci, J.~Henriksson, B.~McPeak, S.~Ricossa, F.~Riva, and A.~Vichi, ``{Positivity bounds on massive vectors},'' \href{https://dx.doi.org/10.1007/JHEP12(2024)051}{{\em JHEP} {\bfseries 12} (2024) 051}, \href{https://arxiv.org/abs/2402.13327}{{\ttfamily arXiv:2402.13327 [hep-th]}}.

\bibitem{Albert:2024yap}
J.~Albert, W.~Knop, and L.~Rastelli, ``{Where is tree-level string theory?},'' \href{https://dx.doi.org/10.1007/JHEP02(2025)157}{{\em JHEP} {\bfseries 02} (2025) 157}, \href{https://arxiv.org/abs/2406.12959}{{\ttfamily arXiv:2406.12959 [hep-th]}}.

\bibitem{Xu:2024iao}
H.~Xu, D.-Y. Hong, Z.-H. Wang, and S.-Y. Zhou, ``{Positivity bounds on parity-violating scalar-tensor EFTs},'' \href{https://dx.doi.org/10.1088/1475-7516/2025/01/102}{{\em JCAP} {\bfseries 01} (2025) 102}, \href{https://arxiv.org/abs/2410.09794}{{\ttfamily arXiv:2410.09794 [hep-th]}}.

\bibitem{Berman:2024kdh}
J.~Berman, ``{Analytic bounds on the spectrum of crossing symmetric S-matrices},'' \href{https://dx.doi.org/10.1007/JHEP08(2025)066}{{\em JHEP} {\bfseries 08} (2025) 066}, \href{https://arxiv.org/abs/2410.01914}{{\ttfamily arXiv:2410.01914 [hep-th]}}.

\bibitem{Haring:2024wyz}
K.~H{\"a}ring and A.~Zhiboedov, ``{What is the graviton pole made of?},'' \href{https://arxiv.org/abs/2410.21499}{{\ttfamily arXiv:2410.21499 [hep-th]}}.

\bibitem{Dong:2024omo}
Z.-Y. Dong, T.~Ma, A.~Pomarol, and F.~Sciotti, ``{Bootstrapping the chiral-gravitational anomaly},'' \href{https://dx.doi.org/10.1007/JHEP05(2025)114}{{\em JHEP} {\bfseries 05} (2025) 114}, \href{https://arxiv.org/abs/2411.14422}{{\ttfamily arXiv:2411.14422 [hep-th]}}.

\bibitem{Beadle:2025cdx}
C.~Beadle, G.~Isabella, D.~Perrone, S.~Ricossa, F.~Riva, and F.~Serra, ``{The EFT bootstrap at finite M$_{PL}$},'' \href{https://dx.doi.org/10.1007/JHEP06(2025)209}{{\em JHEP} {\bfseries 06} (2025) 209}, \href{https://arxiv.org/abs/2501.18465}{{\ttfamily arXiv:2501.18465 [hep-th]}}.

\bibitem{Chang:2025cxc}
C.-H. Chang and J.~Parra-Martinez, ``{Graviton loops and negativity},'' \href{https://dx.doi.org/10.1007/JHEP08(2025)175}{{\em JHEP} {\bfseries 08} (2025) 175}, \href{https://arxiv.org/abs/2501.17949}{{\ttfamily arXiv:2501.17949 [hep-th]}}.

\bibitem{Pasiecznik:2025eqc}
C.~Pasiecznik, ``{Bootstrapping Gravity with Crossing Symmetric Dispersion Relations},'' \href{https://arxiv.org/abs/2506.09884}{{\ttfamily arXiv:2506.09884 [hep-th]}}.

\bibitem{Huang:2025icl}
Y.-t. Huang, S.~Ricossa, F.~Riva, and J.-D. Tsai, ``{The Rise of Linear Trajectories},'' \href{https://arxiv.org/abs/2510.07991}{{\ttfamily arXiv:2510.07991 [hep-th]}}.

\bibitem{Camanho:2014apa}
X.~O. Camanho, J.~D. Edelstein, J.~Maldacena, and A.~Zhiboedov, ``{Causality Constraints on Corrections to the Graviton Three-Point Coupling},'' \href{https://dx.doi.org/10.1007/JHEP02(2016)020}{{\em JHEP} {\bfseries 02} (2016) 020}, \href{https://arxiv.org/abs/1407.5597}{{\ttfamily arXiv:1407.5597 [hep-th]}}.

\bibitem{Aharony:1999ti}
O.~Aharony, S.~S. Gubser, J.~M. Maldacena, H.~Ooguri, and Y.~Oz, ``{Large N field theories, string theory and gravity},'' \href{https://dx.doi.org/10.1016/S0370-1573(99)00083-6}{{\em Phys. Rept.} {\bfseries 323} (2000) 183--386}, \href{https://arxiv.org/abs/hep-th/9905111}{{\ttfamily arXiv:hep-th/9905111}}.

\bibitem{Cremonini:2009sy}
S.~Cremonini, K.~Hanaki, J.~T. Liu, and P.~Szepietowski, ``{Higher derivative effects on eta/s at finite chemical potential},'' \href{https://dx.doi.org/10.1103/PhysRevD.80.025002}{{\em Phys. Rev. D} {\bfseries 80} (2009) 025002}, \href{https://arxiv.org/abs/0903.3244}{{\ttfamily arXiv:0903.3244 [hep-th]}}.

\bibitem{Delsate:2011qp}
T.~Delsate, V.~Cardoso, and P.~Pani, ``{Anti de Sitter black holes and branes in dynamical Chern-Simons gravity: perturbations, stability and the hydrodynamic modes},'' \href{https://dx.doi.org/10.1007/JHEP06(2011)055}{{\em JHEP} {\bfseries 06} (2011) 055}, \href{https://arxiv.org/abs/1103.5756}{{\ttfamily arXiv:1103.5756 [hep-th]}}.

\bibitem{Das:2005za}
S.~R. Das, S.~Giusto, S.~D. Mathur, Y.~Srivastava, X.~Wu, and C.~Zhou, ``{Branes wrapping black holes},'' \href{https://dx.doi.org/10.1016/j.nuclphysb.2005.11.011}{{\em Nucl. Phys. B} {\bfseries 733} (2006) 297--333}, \href{https://arxiv.org/abs/hep-th/0507080}{{\ttfamily arXiv:hep-th/0507080}}.

\bibitem{Donos:2012wi}
A.~Donos and J.~P. Gauntlett, ``{Black holes dual to helical current phases},'' \href{https://dx.doi.org/10.1103/PhysRevD.86.064010}{{\em Phys. Rev. D} {\bfseries 86} (2012) 064010}, \href{https://arxiv.org/abs/1204.1734}{{\ttfamily arXiv:1204.1734 [hep-th]}}.

\bibitem{Cai:2012mg}
R.-G. Cai, T.-J. Li, Y.-H. Qi, and Y.-L. Zhang, ``{Incompressible Navier-Stokes Equations from Einstein Gravity with Chern-Simons Term},'' \href{https://dx.doi.org/10.1103/PhysRevD.86.086008}{{\em Phys. Rev. D} {\bfseries 86} (2012) 086008}, \href{https://arxiv.org/abs/1208.0658}{{\ttfamily arXiv:1208.0658 [hep-th]}}.

\bibitem{Megias:2013joa}
E.~Megias and F.~Pena-Benitez, ``{Holographic Gravitational Anomaly in First and Second Order Hydrodynamics},'' \href{https://dx.doi.org/10.1007/JHEP05(2013)115}{{\em JHEP} {\bfseries 05} (2013) 115}, \href{https://arxiv.org/abs/1304.5529}{{\ttfamily arXiv:1304.5529 [hep-th]}}.

\bibitem{Azeyanagi:2015gqa}
T.~Azeyanagi, R.~Loganayagam, and G.~S. Ng, ``{Anomalies, Chern-Simons Terms and Black Hole Entropy},'' \href{https://dx.doi.org/10.1007/JHEP09(2015)121}{{\em JHEP} {\bfseries 09} (2015) 121}, \href{https://arxiv.org/abs/1505.02816}{{\ttfamily arXiv:1505.02816 [hep-th]}}.

\bibitem{Bhattacharyya:2016knk}
A.~Bhattacharyya, L.~Cheng, and L.-Y. Hung, ``{Relative Entropy, Mixed Gauge-Gravitational Anomaly and Causality},'' \href{https://dx.doi.org/10.1007/JHEP07(2016)121}{{\em JHEP} {\bfseries 07} (2016) 121}, \href{https://arxiv.org/abs/1605.02553}{{\ttfamily arXiv:1605.02553 [hep-th]}}.

\bibitem{Liu:2016hqb}
Y.~Liu and F.~Pena-Benitez, ``{Spatially modulated instabilities of holographic gauge-gravitational anomaly},'' \href{https://dx.doi.org/10.1007/JHEP05(2017)111}{{\em JHEP} {\bfseries 05} (2017) 111}, \href{https://arxiv.org/abs/1612.00470}{{\ttfamily arXiv:1612.00470 [hep-th]}}.

\bibitem{Baggioli:2024zfq}
M.~Baggioli, Y.~Bu, and X.~Sun, ``{Chiral anomalous magnetohydrodynamics in action: effective field theory and holography},'' \href{https://dx.doi.org/10.1007/JHEP04(2025)126}{{\em JHEP} {\bfseries 04} (2025) 126}, \href{https://arxiv.org/abs/2412.02361}{{\ttfamily arXiv:2412.02361 [hep-th]}}.

\bibitem{Dixon:2013uaa}
L.~J. Dixon, \href{https://dx.doi.org/10.5170/CERN-2014-008.31}{``{A brief introduction to modern amplitude methods},''} in {\em {Theoretical Advanced Study Institute in Elementary Particle Physics}: {Particle Physics: The Higgs Boson and Beyond}}, pp.~31--67.
\newblock 2014.
\newblock \href{https://arxiv.org/abs/1310.5353}{{\ttfamily arXiv:1310.5353 [hep-ph]}}.

\bibitem{Bellazzini:2021shn}
B.~Bellazzini, G.~Isabella, M.~Lewandowski, and F.~Sgarlata, ``{Gravitational causality and the self-stress of photons},'' \href{https://dx.doi.org/10.1007/JHEP05(2022)154}{{\em JHEP} {\bfseries 05} (2022) 154}, \href{https://arxiv.org/abs/2108.05896}{{\ttfamily arXiv:2108.05896 [hep-th]}}.

\bibitem{Jacob:1959at}
M.~Jacob and G.~C. Wick, ``{On the General Theory of Collisions for Particles with Spin},'' \href{https://dx.doi.org/10.1006/aphy.2000.6022}{{\em Annals Phys.} {\bfseries 7} (1959) 404--428}.

\bibitem{Froissart:1961ux}
M.~Froissart, ``Asymptotic behavior and subtractions in the mandelstam representation,'' \href{https://dx.doi.org/10.1103/PhysRev.123.1053}{{\em Phys. Rev.} {\bfseries 123} (1961) 1053--1057}.

\bibitem{Martin:1965jj}
A.~Martin, ``{Extension of the axiomatic analyticity domain of scattering amplitudes by unitarity. 1.},'' \href{https://dx.doi.org/10.1007/BF02720568}{{\em Nuovo Cim. A} {\bfseries 42} (1965) 930--953}.

\bibitem{Kaplan:2019soo}
J.~Kaplan and S.~Kundu, ``{A Species or Weak-Gravity Bound for Large $N$ Gauge Theories Coupled to Gravity},'' \href{https://dx.doi.org/10.1007/JHEP11(2019)142}{{\em JHEP} {\bfseries 11} (2019) 142}, \href{https://arxiv.org/abs/1904.09294}{{\ttfamily arXiv:1904.09294 [hep-th]}}.

\bibitem{Kaplan:2020ldi}
J.~Kaplan and S.~Kundu, ``{Closed Strings and Weak Gravity from Higher-Spin Causality},'' \href{https://dx.doi.org/10.1007/JHEP02(2021)145}{{\em JHEP} {\bfseries 02} (2021) 145}, \href{https://arxiv.org/abs/2008.05477}{{\ttfamily arXiv:2008.05477 [hep-th]}}.

\bibitem{Kaplan:2020tdz}
J.~Kaplan and S.~Kundu, ``{Causality constraints in large N QCD coupled to gravity},'' \href{https://dx.doi.org/10.1103/PhysRevD.104.L061901}{{\em Phys. Rev. D} {\bfseries 104} no.~6, (2021) L061901}, \href{https://arxiv.org/abs/2009.08460}{{\ttfamily arXiv:2009.08460 [hep-th]}}.

\bibitem{Afkhami-Jeddi:2018apj}
N.~Afkhami-Jeddi, S.~Kundu, and A.~Tajdini, ``{A Bound on Massive Higher Spin Particles},'' \href{https://dx.doi.org/10.1007/JHEP04(2019)056}{{\em JHEP} {\bfseries 04} (2019) 056}, \href{https://arxiv.org/abs/1811.01952}{{\ttfamily arXiv:1811.01952 [hep-th]}}.

\bibitem{Witten:1979vv}
E.~Witten, ``{Current Algebra Theorems for the U(1) Goldstone Boson},'' \href{https://dx.doi.org/10.1016/0550-3213(79)90031-2}{{\em Nucl. Phys. B} {\bfseries 156} (1979) 269--283}.

\bibitem{Veneziano:1979ec}
G.~Veneziano, ``{U(1) Without Instantons},'' \href{https://dx.doi.org/10.1016/0550-3213(79)90332-8}{{\em Nucl. Phys. B} {\bfseries 159} (1979) 213--224}.

\bibitem{tHooft:1973alw}
G.~'t~Hooft, ``{A Planar Diagram Theory for Strong Interactions},'' \href{https://dx.doi.org/10.1016/0550-3213(74)90154-0}{{\em Nucl. Phys. B} {\bfseries 72} (1974) 461}.

\bibitem{Witten:1979kh}
E.~Witten, ``{Baryons in the 1/n Expansion},'' \href{https://dx.doi.org/10.1016/0550-3213(79)90232-3}{{\em Nucl. Phys. B} {\bfseries 160} (1979) 57--115}.

\bibitem{Israel:1966rt}
W.~Israel, ``{Singular hypersurfaces and thin shells in general relativity},'' \href{https://dx.doi.org/10.1007/BF02710419}{{\em Nuovo Cim. B} {\bfseries 44S10} (1966) 1}. [Erratum: Nuovo Cim.B 48, 463 (1967)].

\bibitem{Barbieri:2003pr}
R.~Barbieri, A.~Pomarol, and R.~Rattazzi, ``{Weakly coupled Higgsless theories and precision electroweak tests},'' \href{https://dx.doi.org/10.1016/j.physletb.2004.04.005}{{\em Phys. Lett. B} {\bfseries 591} (2004) 141--149}, \href{https://arxiv.org/abs/hep-ph/0310285}{{\ttfamily arXiv:hep-ph/0310285}}.

\bibitem{Manohar:1983md}
A.~Manohar and H.~Georgi, ``{Chiral Quarks and the Nonrelativistic Quark Model},'' \href{https://dx.doi.org/10.1016/0550-3213(84)90231-1}{{\em Nucl. Phys. B} {\bfseries 234} (1984) 189--212}.

\bibitem{Randall:1999ee}
L.~Randall and R.~Sundrum, ``{A Large mass hierarchy from a small extra dimension},'' \href{https://dx.doi.org/10.1103/PhysRevLett.83.3370}{{\em Phys. Rev. Lett.} {\bfseries 83} (1999) 3370--3373}, \href{https://arxiv.org/abs/hep-ph/9905221}{{\ttfamily arXiv:hep-ph/9905221}}.

\bibitem{Arkani-Hamed:2000ijo}
N.~Arkani-Hamed, M.~Porrati, and L.~Randall, ``{Holography and phenomenology},'' \href{https://dx.doi.org/10.1088/1126-6708/2001/08/017}{{\em JHEP} {\bfseries 08} (2001) 017}, \href{https://arxiv.org/abs/hep-th/0012148}{{\ttfamily arXiv:hep-th/0012148}}.

\bibitem{Kruczenski:2003be}
M.~Kruczenski, D.~Mateos, R.~C. Myers, and D.~J. Winters, ``{Meson spectroscopy in AdS / CFT with flavor},'' \href{https://dx.doi.org/10.1088/1126-6708/2003/07/049}{{\em JHEP} {\bfseries 07} (2003) 049}, \href{https://arxiv.org/abs/hep-th/0304032}{{\ttfamily arXiv:hep-th/0304032}}.

\bibitem{Afkhami-Jeddi:2018own}
N.~Afkhami-Jeddi, S.~Kundu, and A.~Tajdini, ``{A Conformal Collider for Holographic CFTs},'' \href{https://dx.doi.org/10.1007/JHEP10(2018)156}{{\em JHEP} {\bfseries 10} (2018) 156}, \href{https://arxiv.org/abs/1805.07393}{{\ttfamily arXiv:1805.07393 [hep-th]}}.

\bibitem{Stairs:2002cw}
I.~H. Stairs, S.~E. Thorsett, J.~H. Taylor, and A.~Wolszczan, ``{Studies of the relativistic binary pulsar psr b1534+12: I. timing analysis},'' \href{https://dx.doi.org/10.1086/344157}{{\em Astrophys. J.} {\bfseries 581} (2002) 501--508}, \href{https://arxiv.org/abs/astro-ph/0208357}{{\ttfamily arXiv:astro-ph/0208357}}.

\bibitem{Stairs:1999zr}
I.~H. Stairs, D.~J. Nice, S.~E. Thorsett, and J.~H. Taylor, ``{Recent arecibo timing of the relativistic binary PSR B1534+12},'' in {\em {34th Rencontres de Moriond: Gravitational Waves and Experimental Gravity}}, pp.~309--318.
\newblock 3, 1999.
\newblock \href{https://arxiv.org/abs/astro-ph/9903289}{{\ttfamily arXiv:astro-ph/9903289}}.

\bibitem{Prasanna:2003ix}
A.~R. Prasanna and S.~Mohanty, ``{Constraints on nonminimally coupled curved space electrodynamics from astrophysical observations},'' \href{https://dx.doi.org/10.1088/0264-9381/20/14/304}{{\em Class. Quant. Grav.} {\bfseries 20} (2003) 3023--3028}, \href{https://arxiv.org/abs/gr-qc/0306021}{{\ttfamily arXiv:gr-qc/0306021}}.

\bibitem{Wu:2017yjl}
X.-F. Wu, J.-J. Wei, M.-X. Lan, H.~Gao, Z.-G. Dai, and P.~M{\'e}sz{\'a}ros, ``{New test of weak equivalence principle using polarized light from astrophysical events},'' \href{https://dx.doi.org/10.1103/PhysRevD.95.103004}{{\em Phys. Rev. D} {\bfseries 95} no.~10, (2017) 103004}, \href{https://arxiv.org/abs/1703.09935}{{\ttfamily arXiv:1703.09935 [astro-ph.HE]}}.

\bibitem{Jana:2021lqe}
S.~Jana and S.~Shankaranarayanan, ``{Non-Minimally Coupled Electromagnetic Fields and Observable Implications for Primordial Black Holes},'' \href{https://dx.doi.org/10.3390/universe10070270}{{\em Universe} {\bfseries 10} no.~7, (2024) 270}, \href{https://arxiv.org/abs/2110.06056}{{\ttfamily arXiv:2110.06056 [gr-qc]}}.

\bibitem{Carballo-Rubio:2025zwz}
R.~Carballo-Rubio, H.~Delaporte, A.~Eichhorn, and P.~G.~S. Fernandes, ``{Non-minimal light-curvature couplings and black-hole imaging},'' \href{https://arxiv.org/abs/2505.21431}{{\ttfamily arXiv:2505.21431 [astro-ph.HE]}}.

\bibitem{Chen:2021lvo}
Y.~Chen, Y.~Liu, R.-S. Lu, Y.~Mizuno, J.~Shu, X.~Xue, Q.~Yuan, and Y.~Zhao, ``{Stringent axion constraints with Event Horizon Telescope polarimetric measurements of M87$^{⋆}$},'' \href{https://dx.doi.org/10.1038/s41550-022-01620-3}{{\em Nature Astron.} {\bfseries 6} no.~5, (2022) 592--598}, \href{https://arxiv.org/abs/2105.04572}{{\ttfamily arXiv:2105.04572 [hep-ph]}}.

\bibitem{EliasMiro:2025rqo}
J.~Elias~Mir{\'o}, A.~Guerrieri, M.~A. G{\"u}m{\"u}s, and A.~Zahed, ``{A Geometric View on Crossing-Symmetric Dispersion Relations},'' \href{https://arxiv.org/abs/2509.14170}{{\ttfamily arXiv:2509.14170 [hep-th]}}.

\bibitem{Bellazzini:2025shd}
B.~Bellazzini, A.~Pomarol, M.~Romano, and F.~Sciotti, ``{(Super)$\,$Gravity from Positivity},'' \href{https://arxiv.org/abs/2507.12535}{{\ttfamily arXiv:2507.12535 [hep-th]}}.

\bibitem{BleisteinHandelsman1986}
N.~Bleistein and R.~A. Handelsman, {\em Asymptotic Expansions of Integrals}.
\newblock Dover Publications, 1986.

\bibitem{SteinShakarchi2005}
E.~M. Stein and R.~Shakarchi, {\em Real Analysis: Measure Theory, Integration, and Hilbert Spaces}.
\newblock Princeton Lectures in Analysis, Volume 3. Princeton University Press, 2005.

\bibitem{Rose:1995}
M.~E. Rose, {\em Elementary Theory of Angular Momentum}.
\newblock Dover Publications, New York, dover ed.~ed., 1995.

\end{thebibliography}\endgroup

\end{document}